\documentstyle[aaspp,12pt,epsf]{article}

\def\plotfiddle#1#2#3#4#5#6#7{\vbox to #2{
   \includegraphics{#1}}}

\begin{document}

\title{A Near Infrared Spectroscopic Survey of LINER Galaxies}
\author{J. E. LARKIN}
\affil{Dept. of Astronomy and Astrophysics, University of Chicago and}
\affil{Palomar Observatory, California Institute of Technology}
\authoraddr{Dept. of Astronomy and Astrophysics, 5640 S. Ellis,
Chicago, IL. 60521}
\author{L. ARMUS, R. A. KNOP, B. T. SOIFER \& K. MATTHEWS}
\affil{Palomar Observatory, California Institute of Technology}
\authoraddr{Caltech 320-47, Pasadena, CA 91125}

%-------------------------

\begin{abstract}
This paper reports the results of a near infrared spectroscopic survey
of LINER galaxies undertaken with a new infrared spectrograph at the
5~m Hale telescope.  The galaxy sample includes 11 LINERs with spectra
covering the [FeII] (1.2567~$\mu$m), Pa$\beta$ (1.2818~$\mu$m), H$_2$
( 1-0 S(1), 2.1218~$\mu$m) and Br$\gamma$ (2.1655~$\mu$m) near
infrared emission lines, and one additional galaxy with only [FeII]
and Pa$\beta$ line coverage.

All of the LINERs with infrared line detections have strong
[FeII] and/or H$_2$ emission, with about half (4 out of 9) having
extremely high ratios ($>$2) of [FeII] to Pa$\beta$.  The strength of
the H$_2$ and [FeII] lines is well correlated with the optical [OI]
line, with many LINERs having higher ratios of [FeII]/Pa$\beta$,
H$_2$/Br$\gamma$ and [OI]/H$\alpha$ than other galaxy
types. The LINERs with the highest [FeII]/Pa$\beta$ ratios (termed
``strong'' [FeII] LINERs) show evidence for recent star
formation. Shocks from compact supernova remnants may enhance the
[FeII] emission in these ``strong'' [FeII] LINERs. The LINERs with
lower [FeII]/Pa$\beta$ ratios (termed ``weak'' [FeII] LINERs) are more
consistent with Seyfert-like activity, including higher ionization
states, some strong x-ray sources and some broad H$\alpha$ detections.
The [FeII] luminosity and the [FeII]/Pa$\beta$ ratio in these objects
are more easily explained by hard x-ray excitation than in the
``strong'' [FeII] LINERs.  These ``weak'' [FeII] LINERs are considered
prime candidates for being low luminosity Seyfert nuclei.

\end{abstract}

\section{Introduction}

LINERs (Low Ionization Nuclear Emission-line Region galaxies) are the
most common and lowest energy examples of active galaxies known
(Heckman 1980).  The main LINER characteristic is unusually strong
forbidden line transitions from low ionization states such as
[OII]($\lambda$=3727~\AA), [NII]($\lambda$=6583~\AA), and
[SII]($\lambda$=6717,6731) relative to lines from higher ionization
states.  The original definition used by Heckman (1980) was
[OII]($\lambda$=3727~\AA) / [OIII]($\lambda$=5007~\AA) $>$ 1 and
[OI]($\lambda$=6300~\AA) / [OIII]($\lambda$=5007~\AA) $>$ 1/3.
Classical LINER galaxies also have much less total energy in the
spectral lines as compared to Seyfert galaxies and other types of
active galactic nuclei (AGN); often down by a factor of a 100 in
comparison to Seyfert's.  The weakness of the spectral lines makes
LINERs difficult to study, particularly when the galaxies often have
strong stellar absorption features.

Fast (v$\sim$100~km~s$^{-1}$) shocks (Heckman, 1980), photoionization
from a power-law source (e.g. Ho, Filippenko, \& Sargent 1993) or from
a cluster of very hot stars (e.g. Terlevich \& Melnick 1985; Shields
1992), especially in very dense environments, can all duplicate the
observed emission line flux ratios in LINERs.  An important aspect of
the star formation models, involves compact supernova remnants which
are confined by the high densities within the nuclear region and which
produce strong shocks.  X-ray heating is particularly attractive for
photoionization because hard x-rays are able to penetrate deeply into
molecular clouds creating large partially ionized regions where low
ionization species will dominate.

An intriguing possible variation on the photoionization models
recently proposed by Eracleous, Livio \& Binette (1995), is that
LINERs have a compact object (probably a black hole) in the nucleus
which periodically disrupts a star during a close orbital approach.
As the stellar material accretes onto the central source, high energy
photons are produced and a Seyfert-like broad line region appears.  As
the material is consumed, the ionizing flux drops and the high
ionization states weaken quickly.  Low ionization lines, however,
remain strong for much longer since the decay time is longer and the
light crossing time in this region is much larger than that of the
broad line region. This ``Duty Cycle Hypothesis'' was motivated by the
observation that about 20\% of LINERs had detectable 2300{\AA}
emission with HST (Maoz et al. 1995).  These UV bright galaxies, have
no other obvious difference from the UV dark LINERs.  Under this
theory, the UV bright galaxies would still have ongoing accretion,
while the others are in the quiescent phase.  In support of this
theory, Eracleous et al. (1995) point to NGC 1097 which was observed
to make a sudden transition from a LINER to a Seyfert 1
(Storchi-Bergmann, Baldwin \& Wilson 1993).

It is also possible that LINERs represent a heterogeneous class of
objects.  Some LINERs may have a central ``monster'' like Seyfert
galaxies, while others have one or more dense clusters of hot young
stars.  Shocks may play a role in enhancing the forbidden lines in
either of these two groups.  Whatever the case, extending the number
of observed spectral diagnostics into the infrared gives greater
leverage on the problem of understanding the ultimate origin of the
LINER phenomena.

A significant number of important infrared spectral features exist
which can provide useful astrophysical insights.  Among these are the
Br$\gamma$ (2.1655$\mu$m) and Pa$\beta$ (1.2818$\mu$m) hydrogen
recombination lines.  These lines trace ionizing photons and can be
directly related to other recombination lines and to the strength of
the UV continuum, while suffering much less from dust extinction than
the Balmer lines. Another series of important infrared lines are the
H$_2$ rotation-vibration transitions which have no analog within the
optical. The strongest H$_2$ line is the 1-0 S(1) transition at 2.1218
$\mu$m.  H$_2$ emission is ubiquitous in starburst and Seyfert
galaxies and is thought to originate in slow shocks, UV fluorescence
and X-ray heating.  The forbidden transitions of singly ionized iron
([FeII]), are also strong in the infrared.  Iron is believed to play
an an important role as a coolant in shocked environments (Nussbaumer
and Storey 1988) but in the general interstellar medium it is often
depleted compared to other elements since most iron is locked up in
dust grains. The strongest near infrared [FeII] line is the
1.2567$\mu$m transition in the J band.  This line is often very strong
in Seyfert galaxies and in most cases is thought to trace faster
shocks ($\sim$100 km sec$^{-1}$) than the H$_2$ lines (Graham et
al. 1990).  [FeII] is also thought to be strong in x-ray heated
environments where dust grains have been evaporated.

In this paper, we report new near infrared spectroscopy of a sample of
bright LINERs, and we attempt to use these data to constrain the
possible excitation mechanisms.

%-----------------------------------------------------------------

\section{Observing Procedures and Data Reductions}

%----

\subsection{Target Selection and Observing Strategy}

The target list began with the original sample of 30 LINERs (including
several ``transition'' objects which have line ratios on the border
between LINERs and Seyferts) from Heckman (1980), which are in the
right ascension range between 8$^h$30$^m$ and 13$^h$30$^m$.  To extend
the target list, additional galaxies were selected from Keel (1983)
and Ho et al. (1993).  Because of bad weather during most of the
spring runs, half of the objects observed with the infrared
spectrograph are from these additional surveys.  In the end 11
classical LINERs were observed in both the J and K bands and 1
additional galaxy had only J-band spectra taken.  Table 1 lists the
galaxies observed, the dates of the observations, and the wavelengths
covered.

\begin{table}[hbtp]
\centering
{\small \begin{tabular}{|lccccccc|}
\hline
\multicolumn{8}{|c|}{ } \\[-12pt]
\multicolumn{8}{|c|}{\bf TABLE 1} \\
\multicolumn{8}{|c|}{\bf Near Infrared Spectroscopy Observing Log} \\[3pt]
\hline
\multicolumn{8}{|c|}{ } \\[-12pt]
\hline
\multicolumn{8}{|c|}{ } \\[-12pt]
\multicolumn{1}{|c}{} & & & cz &
 & Central & Res. & P.A. \\
\multicolumn{1}{|c}{Object} & RA(1950) & Dec(1950) & (km s$^{-1}$) &
Date & $\lambda$ & ($\lambda \over {\Delta \lambda}$) & (deg) \\
\multicolumn{1}{|c}{(1)} & (2) & (3) & (4) & (5) & (6) & (7) & (8) \\[3pt]
\hline
\multicolumn{8}{|c|}{ } \\[-12pt]
NGC 0404 & 01h06m39.3s & +35d27m10s & -43 & 94/08/19 & 1.2567 & 1000 & 128 \\
&&&&& 2.1470 & 820 & 128 \\
NGC 2685 & 08h51m40.7s & +58d55m33s & 877 & 95/03/14 & 1.2683 & 1020 & 305 \\
&&&&& 2.1860 & 815 & 305 \\
NGC 3992 & 11h55m00.8s & +53d39m11s & 1051 & 94/05/22 & 1.2707 & 1020 & 270 \\
&&&&& 2.1532 & 800 & 270 \\
NGC 3998 & 11h55m20.9s & +55d43m56s & 1138 & 95/05/13 & 1.2671 & 1020 & 270 \\
&&&&& 2.1762 & 810 & 270 \\
NGC 4258 & 12h16m29.8s & +47d34m51s & 449 & 94/05/22 & 1.2707 & 1020 & 58 \\
&&&&& 2.1532 & 800 & 58 \\
NGC 4589 & 12h35m29.0s & +74d27m59s & 1825 & 95/03/14 & 1.2683 & 1020 & 343 \\
&&&&& 2.1501 & 800 & 343 \\
NGC 4736 & 12h48m31.9s & +41d23m32s & 307 & 95/03/12 & 1.2683 & 1020 & 35 \\
&&&&& 2.1801 & 820 & 35 \\
NGC 4826 & 12h54m16.9s & +21d57m18s & 350 & 95/03/13 & 1.2683 & 1020 & 270 \\
&&&&& 2.1801 & 820 & 270 \\
NGC 5194 & 13h27m46.0s & +47d27m22s & 467 & 94/05/23 & 1.2707 & 1020 & 170 \\
&&&&& 2.1532 & 800 & 170 \\
&&&& 94/07/26 & 1.2837 & 1020 & 170 \\
&&&&& 2.1649 & 810 & 170 \\
NGC 7217 & 22h05m37.9s & +31d06m52s & 1400 & 94/07/26 & 1.2679 & 1020 & 270 \\
NGC 7479 & 23h02m26.41s & +12d03m11s & 2800 & 94/08/21 & 1.2837 & 1060 & 55 \\
&&&&& 2.1649 & 810 & 55 \\
NGC 7743 & 23h41m48.6s & +9d39m25s & 1710 & 94/07/27 & 1.2693 & 1020 & 270 \\
&&&&& 2.1515 & 780 & 270 \\
\hline
\end{tabular}}
\label{t:irlog}
\end{table}

Since LINERs typically have faint emission features, it is important
to concentrate the observations on the potentially brightest near
infrared lines.  Among those lines accessible at zero redshift,
Pa$\beta$ is typically the brightest feature in AGN and starburst
galaxies.  Under case B recombination, Br$\gamma$ is fainter than
Pa$\beta$ by a factor of about six, but is still usually one of the
brightest lines.  The brightest Fe$^+$ line is at 1.2567 $\mu$m which
is only 0.0251 $\mu$m away from Pa$\beta$, making this an obvious pair
of lines to observe simultaneously.  The strongest H$_2$ line
available at zero redshift is from the 1-0 S(1) transition at 2.1218
$\mu$m.  This line is separated by 0.0437 $\mu$m from Br$\gamma$,
again creating an obvious pair of lines to observe simultaneously.  In
order to maximize the number of LINERs observed, only these two
settings were observed.

%-----------------------------------------------------------------

\subsection{Near Infrared Spectroscopy at the 200 Inch Hale Telescope}

Near infrared spectra were obtained using the 200 inch Hale telescope
at Palomar Observatory from 1994 April to 1995 May. The spectra were
taken with a new near infrared longslit spectrometer, described in
Larkin et al. (1996).  For all observations, the lower resolution
grating (R$\sim$1000) was used with the slit set at 0.75'' x 40''.  To
center the objects on the slit, images were first taken through a wide
open slit (10'' x 40'') and the telescope was moved to place the
nuclear centroid onto the slit center.  The position angle of the slit
was selected either by aligning the slit along a position angle of
known extended H$\alpha$ emission (Larkin et al. 1997, in preparation)
or by orienting the slit along the minor axis.  For all objects the
galaxy was moved back and forth along the slit between two locations
10'' from the opposite ends of the slit (20'' between the two
positions). In this way, each spectrum in a pair could serve as the
sky frame for image subtraction of the other mated spectrum.  Two
grating angles were typically used to give rest frame wavelength
coverages from 1.245 to 1.300~$\mu$m (R$\sim$1050), and 2.087 to
2.207~$\mu$m (R$\sim$850). These wavelength ranges were selected to
include the redshifted [FeII] ($\lambda$ = 1.2567$\mu$m), and
Pa$\beta$ ($\lambda$ = 1.2818$\mu$m) spectral lines, and the
redshifted H$_2$ ($\lambda$ = 2.1218$\mu$m) and Br$\gamma$ ($\lambda$
= 2.1655$\mu$m) lines.  All wavelengths quoted in this paper are
values in air.

The Yale Bright Star Catalog (Hoffleit 1964) was used to select main
sequence G stars that were observed close in time and airmass to the
galaxies in order to remove telluric absorption from the galaxy
spectra and to serve as flat fields.  The chopping secondary of the
telescope was driven with a triangular waveform to move the star back
and forth along the slit to uniformly illuminate the array.  The stars
represent a good approximation to an ideal blackbody at these
wavelengths except for weak hydrogen absorption lines. In order to
properly correct the galactic spectra, these lines must first be
removed from the stellar spectra by interpolation.  For all of the
galaxies in the sample except NGC~404, the redshift is high enough
($>$300 km s$^{-1}$) that contamination from the interpolated portion
of the G star spectra should not affect the emission lines observed.

The galaxy spectra were reduced by first subtracting image pairs.  The
galaxy frames were then divided by the interpolated star frames.  Bad
pixels were then replaced by the median of their neighboring
pixels. Atmospheric OH emission lines in the stellar spectrum were fit
with third order polynomials in order to remove a slight curvature in
the spectral lines and to wavelength calibrate the spectra.
Wavelengths for the OH lines were obtained from Oliva and
Origlia~(1992).  Slight spatial curvature was removed by fitting
spectra taken of a calibrator star which was moved in 5'' steps along
the slit.

%----

\subsection{Near Infrared Imaging at the 200 Inch Hale Telescope}

Broad band 1.25$\mu$m (J-band) and 2.2$\mu$m (K-band) images taken
with a mirror placed in front of the grating were used to flux
calibrate the spectra.  These images, were themselves flux calibrated
by matching synthetic aperture fluxes in the images with aperture
photometry from either Neugebauer et al. (1987), Aaronson (1977), or
Willner et al. (1985). A 0.75'' x 3'' synthetic aperture on the same
broad band images was then used to determine the flux densities within
the slit for the spectroscopic observations. These ``slit''
measurements were then used to scale the counts in the spectra to
actual flux.

%----

%-----------------------------------------------------------------

\section{Results}

The flux calibrated spectra for the 12 classical LINERs observed at
$\sim$1000 resolution are shown in figure 1.  The J-band
spectra are shown in the left panel and the K-band in the right.  Many
of the spectra have been shifted upwards to prevent overlap and the
amount of the shift is indicated in parenthesis next to the galaxy's
name. The locations of the most common emission and absorption
features are marked at the top of each panel.  The [FeII] line at
1.2567$\mu$m is the most common emission line detected, being present
in 8 of the 12 galaxies.  The H$_2$ line at 2.1218$\mu$m is also a
common feature, being found in 4 of the 11 galaxies with K-Band
spectra.  Pa$\beta$ (1.2818$\mu$m), which is typically the strongest
near infrared line available at zero redshift, is only found in
emission in 2 galaxies, NGC~5194 and NGC~7479.  Br$\gamma$
(2.166$\mu$m) is undetected in all 11 galaxies with K-Band spectra
(note some of the spectra show features near Br$\gamma$ due to
residual hydrogen absorption features in the calibration stars).
Detection limits varied between the objects but lines as faint as 1.0
$\times$ 10$^{-15}$ ergs s$^{-1}$ cm$^{-2}$ were detected in some
galaxies.  Table 2 gives the measured fluxes and uncertainties for the
infrared lines. In most cases a 3'' by 1000 km s$^{-1}$ aperture was
placed on the continuum subtracted spectra to determine the fluxes
making the true aperture on the sky 3'' by 0.75'' (the slit width).
The uncertainties were determined by placing the same aperture at
several locations on the spectra above and below each emission
line. In many cases, Pa$\beta$ is detected in absorption, and the
strength is listed as a negative value in the table.  Table 2 also
gives the equivalent widths of the [FeII] and H$_2$ lines in columns 6
and 7.  In all cases, the emission lines were not resolved spatially
with seeing of $\sim$1.0 arcsec.

%-----------------------------------------------------------------
%-----------------------------------------------------------------

\begin{table}[hbtp]
\centering
\begin{tabular}{|lcccccc|}
\hline
\multicolumn{7}{|c|}{ } \\[-12pt]
\multicolumn{7}{|c|}{\bf TABLE 2} \\
\multicolumn{7}{|c|}{\bf Measured Line Fluxes and Equivalent widths } \\
\hline
\multicolumn{7}{|c|}{ } \\[-12pt]
\hline
\multicolumn{7}{|c|}{ } \\[-12pt]
\multicolumn{1}{|c}{} & \multicolumn{4}{c}{flux (x10$^{-15}$ erg cm$^{-2}$
s$^{-1}$)} & \multicolumn{2}{c|}{EqW. (\AA)} \\
\multicolumn{1}{|c}{Object} & [FeII] & Pa$\beta$ & H$_2$ & Br$\gamma$ &
EqW([FeII]) & EqW(H$_2$) \\
\multicolumn{1}{|c}{(1)} & (2) & (3) & (4) & (5) & (6) & (7) \\
\hline
\multicolumn{7}{|c|}{ } \\[-12pt]
NGC 0404 & 7.8$\pm$0.2 & -1.3$\pm$0.3 & 2.4$\pm$0.7 & 0$\pm$1$^b$ &
3.1$\pm$0.3 & 3.1$\pm$0.3 \\
NGC 2685 & 0$\pm$1 & -1.5$\pm$1 & 0$\pm$1 & 0$\pm$1 & $<$0.5 & $<$2 \\
NGC 3992 & 0$\pm$1 & -2$\pm$1 & 0$\pm$0.5 & 0$\pm$0.5 & $<$0.7 & $<$1 \\[8pt]
NGC 3998 & 3.1$\pm$0.7 & -1$\pm$1 & 0$\pm$1 & 0$\pm$1 & 1.7$\pm$0.2 & $<$1 \\
NGC 4258 & 3.5$\pm$0.6 & -0.4$\pm$0.6 & 0$\pm$0.9 & 0$\pm$0.9 & 1.5$\pm$0.2 &
$<$1 \\
NGC 4589 & 0$\pm$2 & -2.6$\pm$2 & 0$\pm$0.8 & 0$\pm$0.8 & $<$1 & $<$1 \\[8pt]
NGC 4736 & 10$\pm$4 & -8$\pm$3 & 0$\pm$2 & 0$\pm$4 & 0.8$\pm$0.4 & $<$0.5 \\
NGC 4826 & 4$\pm$2 & -2.6$\pm$2 & 0$\pm$0.5 & 0$\pm$2 & 0.9$\pm$0.4 & $<$0.3 \\
NGC 5194$^a$ & 5.8$\pm$0.9 & 1.0$\pm$0.7 & 3.9$\pm$0.6 & 0$\pm$1 &
2.8$\pm$0.4 & 5.6$\pm$0.8 \\
& 7.8$\pm$0.9 & 1.5$\pm$0.4 & 4.7$\pm$0.7 & 0.5$\pm$.3 & 3.7$\pm$0.5 &
 7$\pm$1 \\[8pt]
NGC 7217 & 2.3$\pm$0.7 & -1.5$\pm$0.7 & - & - & 1.4$\pm$0.4 & - \\
NGC 7479 & 0$\pm$0.3 & 1.3$\pm$0.4 & 2.8$\pm$0.6 & 0$\pm$0.6$^b$ & $<$0.5 &
9$\pm$2 \\
NGC 7743 & 4.5$\pm$1 & -1.2$\pm$1 & 3.7$\pm$0.4 & 0$\pm$0.5 & 2.8$\pm$0.6 &
4.6$\pm$0.6 \\
\hline
\end{tabular}

\noindent {}

\raggedright
\noindent $^a$ NGC 5194 was observed twice, the May 94 values are given
above the July 94 values.

\noindent $^b$ Although a small bump or wiggle appears near Br$\gamma$ in
these galaxies, it is probably noise or residual from the calibration star.
\label{t:cflux}
\end{table}

% Figure 1
\begin{figure}[p]
\plotfiddle{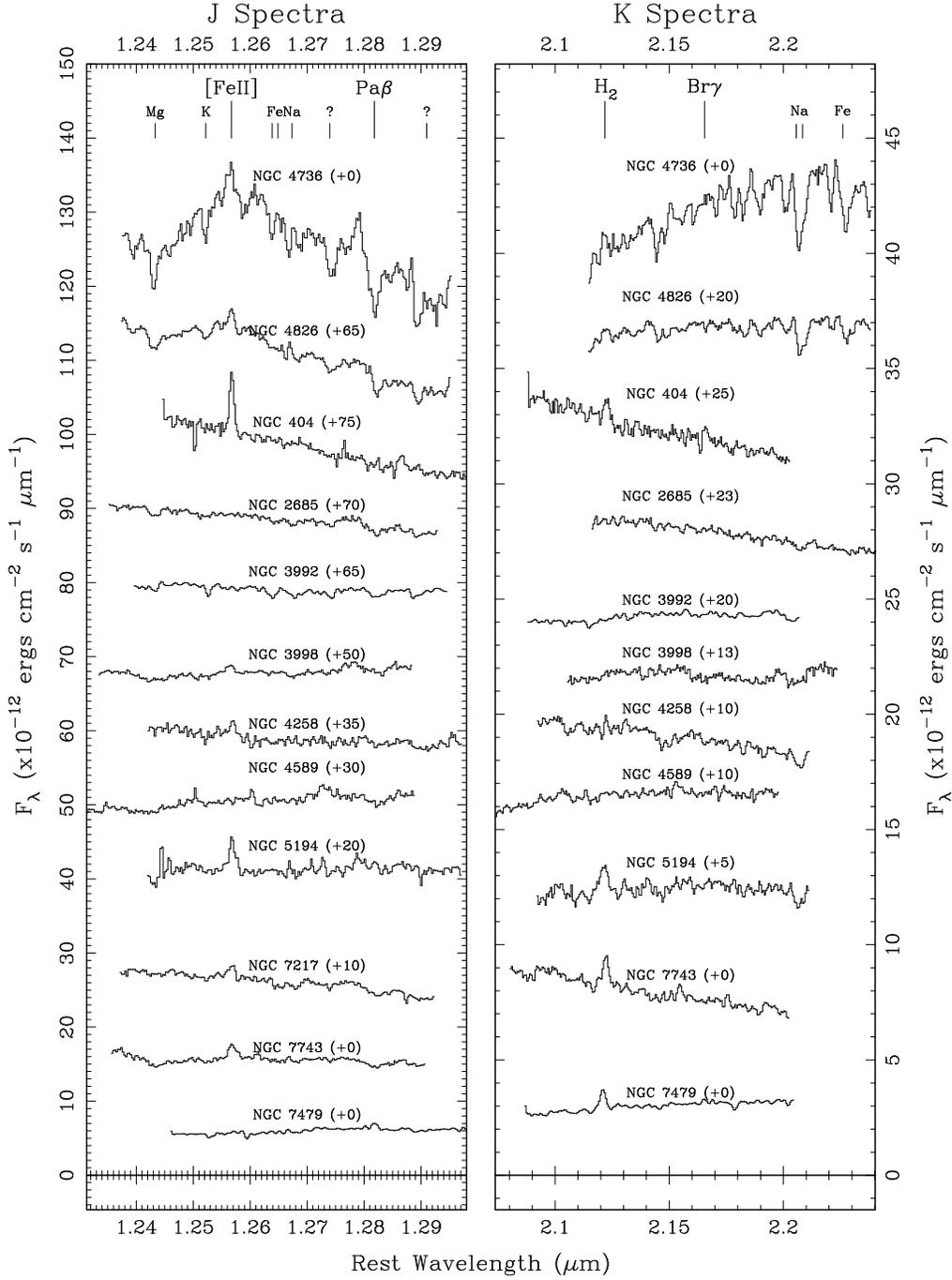}{7.truein}{0}{80}{80}{-20}{-560}
\caption{ Flux calibrated spectra of individual galaxies: The
left spectra are in the J-band window and include the wavelengths of
[FeII] ($\lambda_{rest}$=1.2567$\mu$m) and Pa$\beta$ at a resolution
of $\sim$1050. The right spectra are in the K-band window and include
the wavelengths of H$_2$ ($\lambda_{rest}$=2.1218 $\mu$m) and
Br$\gamma$ at a resolution of $\sim$850.  The spectra of each galaxy
have been shifted up by a constant amount to prevent the spectra from
overlapping.  The amount shifted up is in units of
10$^{-12}$~ergs~cm$^{-2}$~s$^{-1}$~$\mu$m$^{-1}$ and is given in
parenthesis next to each galaxies name. The spectra have also been
shifted in wavelength to the galaxy's restframe. The locations of
emission lines and atomic absorption features are also marked at the
top of each graph.  The emission lines are given in larger type and
are marked higher than the atomic lines.}
\end{figure}
\clearpage

\subsection{Estimated Pa$\beta$ and Br$\gamma$ strengths}

Even though Pa$\beta$ and Br$\gamma$ are usually not detected in
emission, it is important to try and estimate their strengths because
these lines are directly proportional to the ionizing photon flux, and
provide useful reddening insensitive ratios with the [FeII] and H$_2$
lines, respectively.  We use an extrapolation from the optical
recombination lines to estimate the infrared recombination line
emission strengths. This technique uses the case B recombination
ratios for hydrogen (Osterbrock 1989).  Differential reddening between
H$\alpha$ and Pa$\beta$ was removed with the extinction coefficients
of Cardelli et al. (1989).  This technique was also applied to
estimate Br$\gamma$ even though the flux should in most cases be below
our detection limits.  The equations relating H$\alpha$ and H$\beta$
observed line fluxes to P$\beta$ and Br$\gamma$ observed line fluxes
under these conditions are:

\begin{equation}
f(Pa\beta) = 0.0141 \times AC \times f(H\alpha) \times \left(
{f(H\alpha) \over f(H\beta)} \right)^{1.272}
\label{e:pab}
\end{equation}

\begin{equation}
f(Br\gamma) = 0.0016 \times AC \times f(H\alpha) \times \left(
{f(H\alpha) \over f(H\beta)} \right)^{1.615}.
\label{e:brg}
\end{equation}

\noindent
where f(x) is the observed flux of line x.  An aperture correction
factor, AC, has been added to correct for the different apeture sizes
used in the optical and infrared measurements.  If H$\alpha$/H$\beta$
was not known, the median H$\alpha$ to H$\beta$ ratio of 3.81 (A$_V$ =
0.53 mag) from the other galaxies was used. In these cases the
estimated values are given in parenthesis in table 3.  The aperture
correction between our 3'' by 0.75'' slit and the larger apertures in
the other papers depends on the spatial extent of the line emission.
We used H$\alpha$+[NII] images obtained at the 60~inch telescope at
Palomar (Larkin et al. 1997 in preparation), to determine the
appropriate aperture correction for each object for the H$\alpha$ and
H$\beta$ measurements of Heckman et al. (1980), Keel et al.  (1983)
and Ho et al.(1993).  For the larger circular apertures used in
Heckman et al. (1980) and Keel et al. (1983), the average correction
is 0.34$\pm$0.04, while the average corrections to Ho et al.'s (1993)
1''$\times$4'' and 2''$\times$4'' apertures were 0.73$\pm$0.03 and
0.44$\pm$0.05 respectively.  The final estimated Pa$\beta$ and
Br$\gamma$ fluxes are given in columns (4) and (5) respectively of
table 3.  It is important to remember that these predicted fluxes are
for Pa$\beta$ and Br$\gamma$ as they would be measured, not corrected
for extinction.  The only corrections are for the difference in
reddening between the optical and the infrared lines.

%-----------------------------------------------------------------
%-----------------------------------------------------------------

\begin{table}[hbtp]
\centering
\begin{tabular}{|lcccccc|}
\hline
\multicolumn{7}{|c|}{ } \\[-12pt]
\multicolumn{7}{|c|}{\bf  TABLE 3} \\
\multicolumn{7}{|c|}{\bf  Estimated Pa$\beta$ and Br$\gamma$ Fluxes and
Ratios} \\[3pt]
\hline
\multicolumn{7}{|c|}{ } \\[-12pt]
\hline
\multicolumn{7}{|c|}{ } \\[-12pt]
\multicolumn{1}{|c}{} & \multicolumn{4}{c}{flux (x10$^{-15}$ erg cm$^{-2}$
s$^{-1}$)} & & \\
\cline{2-5}
\multicolumn{7}{|c|}{ } \\[-12pt]
\multicolumn{1}{|c}{Object} &
$<$Pa$\beta$ abs.$>$ & 
est. Pa$\beta$ & $<$Pa$\beta>$ & $<$Br$\gamma>$ & $[FeII] \over <Pa\beta>$ &
$H_2 \over <Br\gamma>$ \\
\multicolumn{1}{|c}{(1)} & (2) & (3) & (4) & (5) & (6) & (7) \\[3pt]
\hline
\multicolumn{7}{|c|}{ } \\[-10pt]
NGC 0404 & 2.5$\pm$0.5 & 1.2$\pm$.6 & 3.1$^b$,2.7$^c$ & .55$^b$,.49$^c$ &
2.7$\pm$.3 & 5$\pm$2\\
NGC 2685 & 2$\pm$.6 & .5$\pm$1 & ($<$.4)$^a$ & ($<$.06)$^a$ & - & - \\
NGC 3992 & 1.7$\pm$.6 & -.3$\pm$1 & ($<$.3)$^a$ & ($<$.05)$^a$ & - & -
\\[8pt]
NGC 3998 & 2.2$\pm$.7 & 1.2$\pm$1 & 3.6$^a$,7.9$^b$,4.3$^c$ &
.61$^a$,1.5$^b$,.72$^c$ & .6$\pm$.3 & $<2$ \\
NGC 4258 & 2.8$\pm$.9 & 2.4$\pm$1 & (5.7)$^a$ & (1.0)$a$ & .6$\pm$.3 &
$<2$ \\
NGC 4589 & 2.5$\pm$.8 & 0$\pm$2 & ($<$.24)$^a$ & ($<$.04)$^a$  & - & -
\\[8pt]
NGC 4736 & 15$\pm$5 & 7$\pm$6 & (1.9)$^c$ & (.35)$^c$ & 5.3$\pm$2.5 & $<20$\\
NGC 4826 & 5$\pm$1 & 2.5$\pm$2 & (6.0)$^c$ & (1.1)$^c$  & .7$\pm$.4 & $<1$ \\
NGC 5194 & 2.6$\pm$.9 & 3.6$\pm$1 & (3.1)$^b$,(2.7)$^c$ &
(.55)$^b$,(.48)$^c$ & 2.3$\pm$.6 & 8$\pm$3 \\[8pt]
NGC 7217 & 1.8$\pm$.6 & .3$\pm$.8 & .98$^b$ & .16$^b$ & 2.3$\pm$.9 & - \\
NGC 7479 & .7$\pm$.3 & 2$\pm$.5 & 1.2$^b$ & .26$^b$ & $<$.5 & 11$\pm$4 \\
NGC 7743 & 1.8$\pm$.6 & .6$\pm$1 & 3.4$^b$ & .74$^b$ & 1.3$\pm$.4 & 5$\pm2$ \\
\hline
\end{tabular}

\raggedright

\noindent {}

\noindent $^a$ Flux estimated from Heckman et al. (1980).

\noindent $^b$ Flux estimated from Ho et al. (1993).

\noindent $^c$ Flux estimated from Keel et al. (1983).

\raggedright
\label{t:eflux}
\end{table}

%-----------------------------------------------------------------
\clearpage

\subsection{Pa$\beta$ absorption strength}

Since many of the predicted Pa$\beta$ emission line strengths are
within the range of detectability, it is important to understand why
these lines were not seen.  The most likely possibility is that
stellar absorption lines in the galaxies are equal to or stronger than
the emission lines.  In order to evaluate this possibility, we looked
at the three galaxies, NGC~2685, NGC~3992, and NGC~4589, which have
the weakest predicted Pa$\beta$ emission line strengths (all have only
upper limits on their optical H$\alpha$ and H$\beta$ fluxes).  In
these three galaxies, the measured absorption is much larger than the
upper limit on emission, and the emission can be assumed to have
little impact on the measured values. The average equivalent width of
the Pa$\beta$ absorption in these three objects is 1.2$\pm$0.25 \AA,
comparable to that seen in late G type stars.  Assuming that the other
galaxies in this sample have comparable or stronger absorption, column
(2) of table 3 gives the lower limit on Pa$\beta$ absorption for each
galaxy.  Adding this to the measured Pa$\beta$ fluxes in table 2 gives
the lower limit on the true Pa$\beta$ emission line strength and these
values are listed in column (3) of table 3.  For all of the galaxies
with Pa$\beta$ seen in absorption, this fiducial absorption correction
is sufficient to make the corrected values positive or zero,
indicating that absorption is at least comparable to the emission line
strengths, and is a plausible explanation for the nondetections.

Five of the galaxies (NGC~404, NGC~3998, NGC~4258, NGC~4826, and
NGC~7743) have expected Pa$\beta$ fluxes extrapolated from the
measured H$\alpha$ fluxes which are significantly higher than these
absorption corrected values.  This probably indicates that the average
absorption correction is too small, and these galaxies have absorption
strengths closer to that found for A type stars. These galaxies may,
therefore, have younger stellar populations than the three galaxies
used in determining the amount of absorption.  These five galaxies are
among those with strong [FeII] detections.  Although highly
suggestive, it is important to remember that the expected Pa$\beta$
fluxes are extrapolated from H$\alpha$ and H$\beta$ measurements and
unexpected factors such as unusual line ratios or patchy extinction
could be affecting these values.

\subsection{[FeII] and H$_2$ line ratios}

As a first step in the analysis of the [FeII] and H$_2$ emission
lines, we calculate the [FeII]-to-Pa$\beta$ and H$_2$-to-Br$\gamma$
line flux ratios and compare them to other galaxy types in the
literature.  Although these line flux ratios are reddening insensitive
in principle, the Pa$\beta$ and Br$\gamma$ line fluxes are dependent
on the ratio of H$\alpha$ to H$\beta$ used to infer A$_V$.  If the
extinction were very patchy towards the nucleus or had an unusual
wavelength dependence, these line ratios could be biased.  The ratios
of [FeII] to Pa$\beta$ are shown in column (6) of table 3, and the
H$_2$ to Br$\gamma$ ratios are similarly given for each galaxy in
column (7).  These use the extrapolated Pa$\beta$ and Br$\gamma$
values discussed in section 3.1.  In principle, these values can be
treated as ratios to the optical H$\alpha$ fluxes corrected for
extinction, and aperture size and then scaled to the relative
Pa$\beta$ or Br$\gamma$ to H$\alpha$ intensity.  This also allows for
easy comparison with other emission line ratios which use H$\alpha$ or
H$\beta$.

Figure 2 plots [FeII]/Pa$\beta$ versus H$_2$/Br$\gamma$ for the LINERs
and other objects taken from the literature.  It is important to
remember that several of the objects such as Arp~220 and NGC~5128 (Cen
A), although they are marked as LINERs and do satisfy the LINER
spectral criterion, they are very different types of objects from the
``Classical'' LINERs included in this survey.  However, it is also
true that the LINER classification may apply to a diverse group of
objects, and that the relation of objects like Arp 220 and NGC 5128 to
the lower power LINERs may be very enlightening.  Several of the
LINERs have both ratios significantly higher than most other objects.
For all included galaxies, regardless of galaxy type, a linear
correlation is also evident over a range in ratios of more than 100.
The one exception is the LINER NGC~7479 which has weaker [FeII]
emission than other LINERs given its large H$_2$/Br$\gamma$.

% figure 2
\begin{figure}[htbp]
\plotfiddle{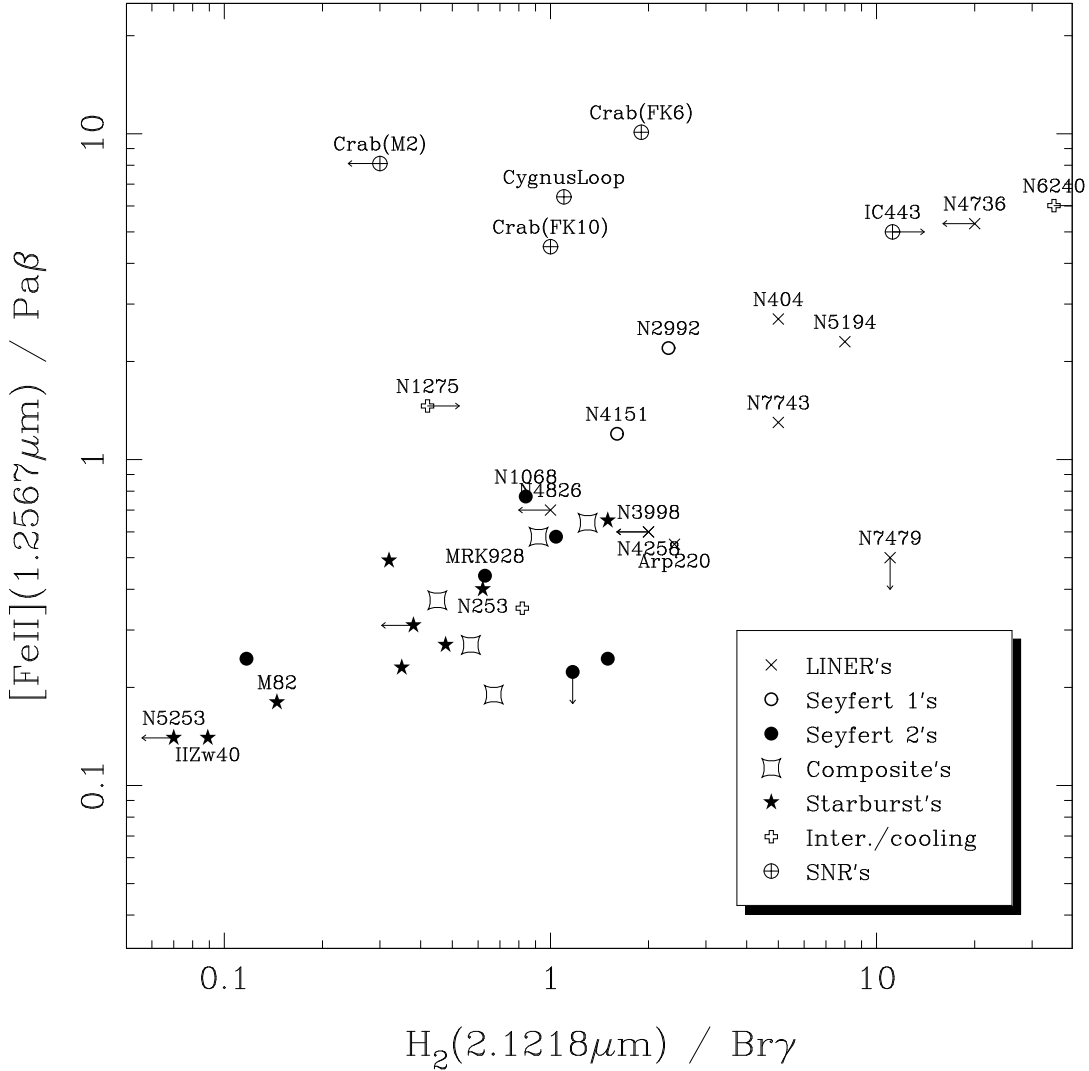}{7truein}{0}{90}{90}{-34}{-650}
\caption{ For all of the objects in this sample plus many
objects from the literature, the ratio [FeII]/Pa$\beta$ is plotted
against the ratio H$_2$/Br$\gamma$. There is a strong linear
correlation between these two ratios for the galaxies included,
regardless of class, even though some of the Seyfert 2's lie
significantly below the correlation.  Several galactic SNR's are also
plotted and usually have unusually strong [FeII]/Pa$\beta$ flux ratios
as compared to galaxies with similar H$_2$/Br$\gamma$ ratios. Also,
NGC~7479, which is often classified as a Seyfert 2, is much weaker in
[FeII] compared to H$_2$ than the correlation predicts.  }
\end{figure}

The points for the supernova remnants (SNR's) in figure 2
are for small apertures on individual filaments and may not represent
the global ratios in these objects.  One of the SNR's in figure
2 is the Cygnus Loop, where the excitation is thought to
arise from an unusually fast shock with a magnetic precursor
penetrating a low density environment (2 cm$^{-3}$)(Graham et
al. 1991). The other SNR points are from various locations in the Crab
Nebula, where Graham et al. (1990) argue photoionization, not shock
heating is the dominant emission source.  The SNR with the largest
H$_2$/Br$\gamma$ is IC~443, where the H$_2$ emission is believed to be
shock excited (Graham et al. 1987).

\clearpage

Figure 3 plots H$_2$/Br$\gamma$ versus OI/H$\alpha$ for all of the
classical LINERs and for many galaxies in the literature.  Mouri et
al. (1989) found a linear correlation for starburst and Seyfert
galaxies between the H$_2$ and OI(6300\AA) lines, and the LINERs in
this sample appear to extend the linear correlation to higher values
by a factor of 2 or more, although there is significant scatter.  For
the galaxies NGC~4258 and NGC~4826 the OI/H$\alpha$ ratio is close to
that found in Seyfert galaxies and the upper limit on H$_2$ emission
also groups the objects with Seyferts.  This is not unexpected since
both are considered prime candidates for dwarf Seyferts (Filippenko
and Sargent 1985).  Seven objects have an H$_2$/Br$\gamma$ ratio above
3, including 6 LINERs and the unusual object NGC~6240 which does
satisfy the optical classification for a LINER.  These are the only
known extragalactic objects with such a large ratio of H$_2$ to
Br$\gamma$.

% figure 3
\begin{figure}[hbtp]
\plotfiddle{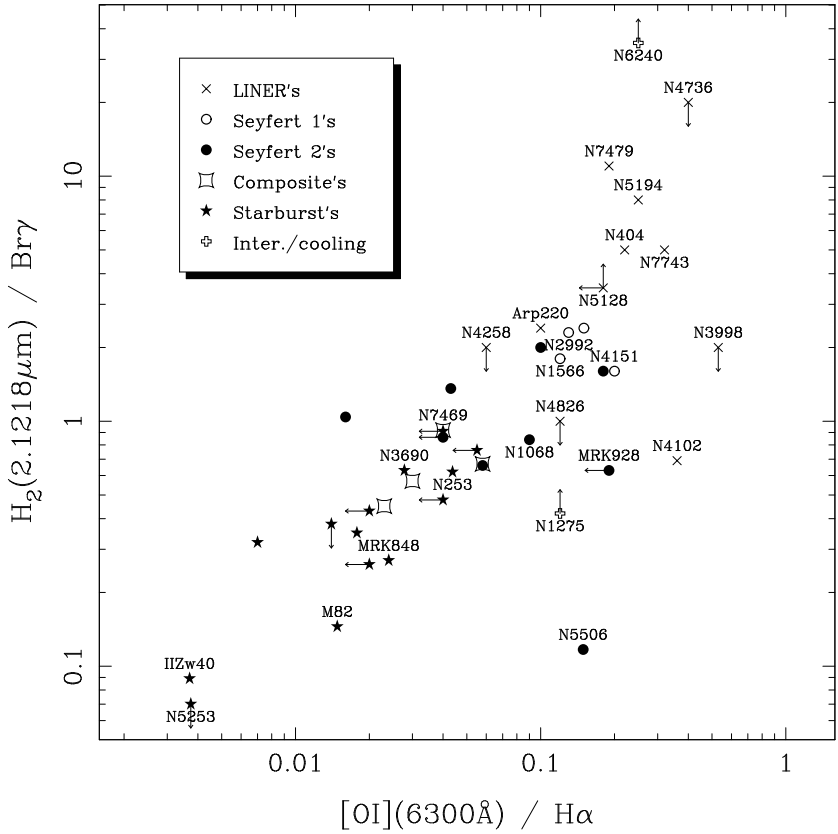}{5truein}{0}{70}{70}{0}{-470}
\caption{For all of the objects in this sample plus all
available objects in the literature, the ratio H$_2$/Br$\gamma$ is
plotted against the ratio OI/H$\alpha$.}
\end{figure}

\clearpage

Figure 4 plots [FeII]/Pa$\beta$ versus OI/H$\alpha$ for all
of the classical LINERs and for many galaxies in the literature.
The linear correlation seen in the previous two figures is still present,
but the scatter is more apparent.  Again, LINERs
in general have both ratios higher than Seyferts or starbursts.

%figure 4
\begin{figure}[hbtp]
\plotfiddle{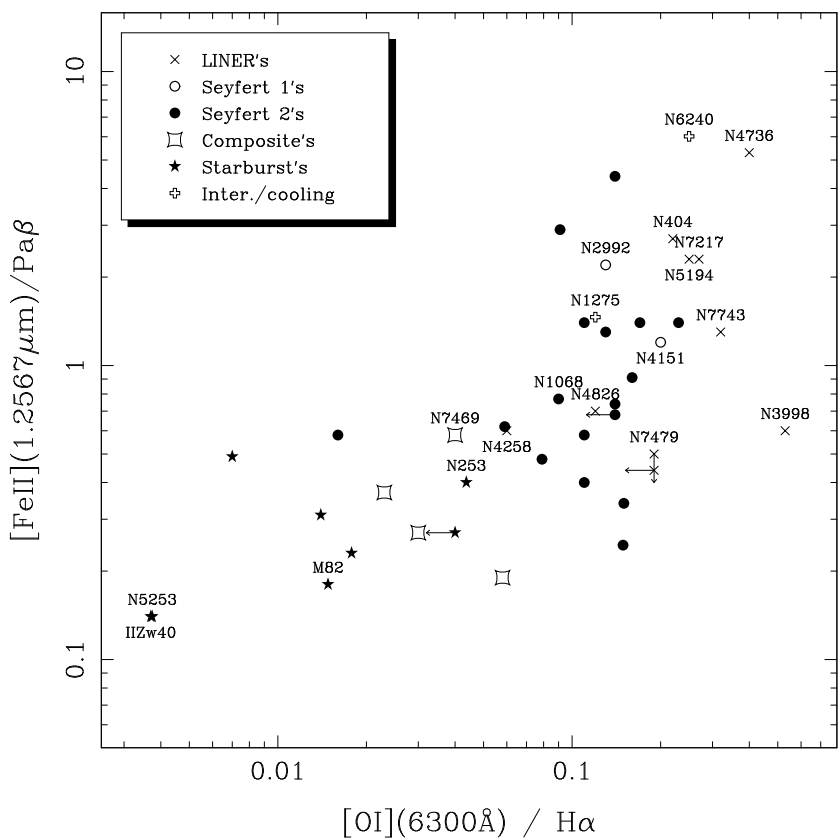}{5truein}{0}{70}{70}{-0}{-470}
\caption{ For all of the objects in this sample plus all
available objects in the literature, the ratio [FeII]/Pa$\beta$ is
plotted against the ratio OI/H$\alpha$.  The correlation between
[FeII] and [OI] is weaker than the correlations between H$_2$ and
[FeII] and between H$_2$ and [OI] shown in figures 2 and
3 }
\end{figure}

\clearpage

Figure 5 plots the optical [OIII]/H$\beta$ ratio against
H$_2$/Br$\gamma$ for the objects in this sample and the available
objects in the literature.  Unlike figures 2-4, no obvious correlation
is evident, however, figure 5 shows a clear separation
between the LINERs and Seyfert galaxies primarily due to the
difference in H$_2$/Br$\gamma$.  As mentioned above, the six known
objects with H$_2$/Br$\gamma$ ratios above 3 have LINER type spectra.
Starburst galaxies are located in the lower left corner of these
figures and are well separated from the other types. The
[FeII]/Pa$\beta$ ratio is significantly poorer at separating the
various galaxy classes. The H$_2$/Br$\gamma$ ratio is probably better
at separating the classes than low ionization optical lines, because
H$_2$ traces the colder molecular gas and is therefore less closely
coupled with the high ionization lines like [OIII]. The three LINERs
which are not grouped with the others in figure 5, all
have unusual activity.  NGC~4258 is a transition object with
substantial recent evidence for a central black hole (see discussion
of individual objects in the appendix). It is not unexpected, then
that it falls within the Seyferts in this figure.  NGC~4826 has an
extrapolated Pa$\beta$ flux almost twice as strong as can be accounted
for with the fiducial absorption applied in section 3.2.  This may
imply a young stellar population and recent star formation, and could
explain why it is consistent with the starburst galaxies in the
figure.  NGC~3998, as argued below has a central source with
Seyfert-like ratios, while in large apertures (D$\sim$5'') the low
ionization lines are stronger and it appears to be a true LINER.  This
could explain why our small aperture shows such a low H$_2$/Br$\gamma$
ratio, and why it lies within the range of Seyfert values.

% figure 5
\begin{figure}[hbtp]
\plotfiddle{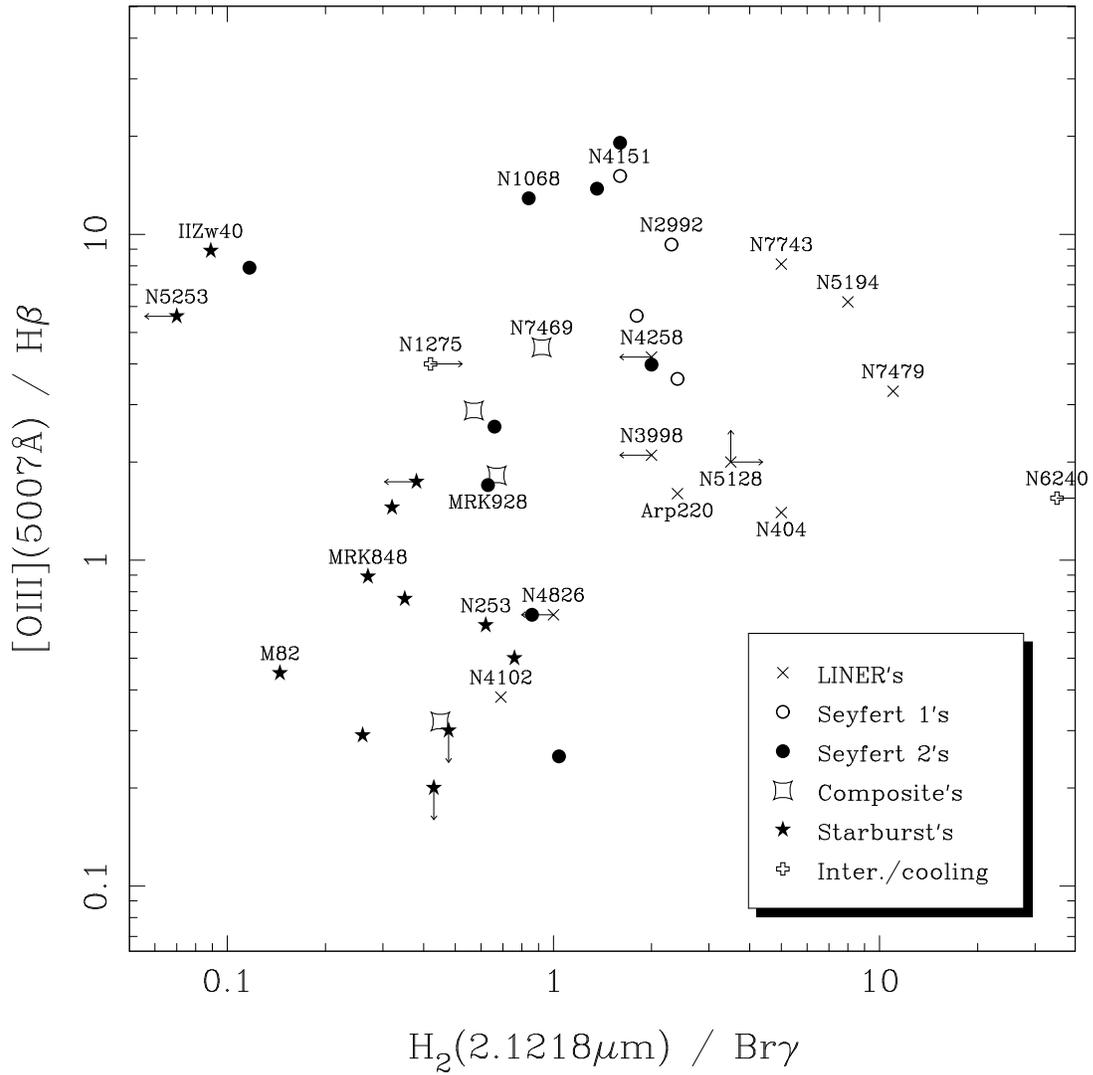}{7truein}{0}{90}{90}{-34}{-650}
\caption{ For all of the objects in this sample plus objects in
the literature, the ratio [OIII]/H$\beta$ is plotted against the ratio
H$_2$/Br$\gamma$. }
\end{figure}

\clearpage

\subsection{Comparison with IRAS $\alpha$(25:60) index}

In figures 6 and 7, the IRAS 25 to 60 micron index is plotted against
the H$_2$/Br$\gamma$ and [OI]/H$\alpha$, respectively for the LINERs
in our sample and from the literature as well as many starbursts,
Seyferts, composites (Seyfert + starburst), and Ultraluminous Infrared
Galaxies (ULIRGS) also from the literature.  The index is defined as:

\begin{equation}
\alpha(25:60) = {{ - log( f_{25\mu m} / f_{60\mu m} ) } \over {
log( 25\mu m / 60\mu m ) }}.
\label{e:2560}
\end{equation}

\noindent
The IRAS indices measure the shape of the far infrared continuum and
$\alpha$(25:60) in particular is thought to reflect the relative
importance of warm (30-50 K) dust which dominates the 60$\mu$m band to
hot (100-150 K) dust which dominates the 25$\mu$m band.  For
reference, non-active galaxies have a large range of $\alpha$(25:60),
but are typically between -2 and -3.  Mouri and Taniguchi (1992) found
that starburst galaxies and to a lesser extent, infrared ultraluminous
galaxies, showed a linear correlation between this IRAS index and
H$_2$/Br$\gamma$.  They further found that Seyfert galaxies showed no
such correlation, and although some Seyferts have $\alpha$(25:60)
comparable to starbursts, most Seyferts had a shallower 25 to 60
$\mu$m slope than the starburst correlation predicts.  In fact, the
$\alpha$(25:60) index has been found to be an efficient identifier of
Seyfert galaxies, especially the most luminous objects where the host
galaxy colors play less of a role (De Grijp, et al. 1992).  The LINERs
show a large spread in $\alpha$(25:60) from NGC~3998 at -1.15 to
NGC~4826 at -3.23, while showing little variation in H$_2$, and
[OI]. LINERs also show a weak trend that the objects with flatter
$\alpha$(25:60) have weaker [FeII] than the steeper index LINERs.
Some of the LINERs and Seyferts are close to the starburst
correlations, but most have flatter indices than either the
H$_2$/Br$\gamma$ of [OI]/H$\alpha$ would predict.  ULIRGs show the
opposite trend with more of them on the steeper side of the
relationship, indicating an unusually large ratio of warm to hot dust.

% figure 6
\begin{figure}[hbtp]
\plotfiddle{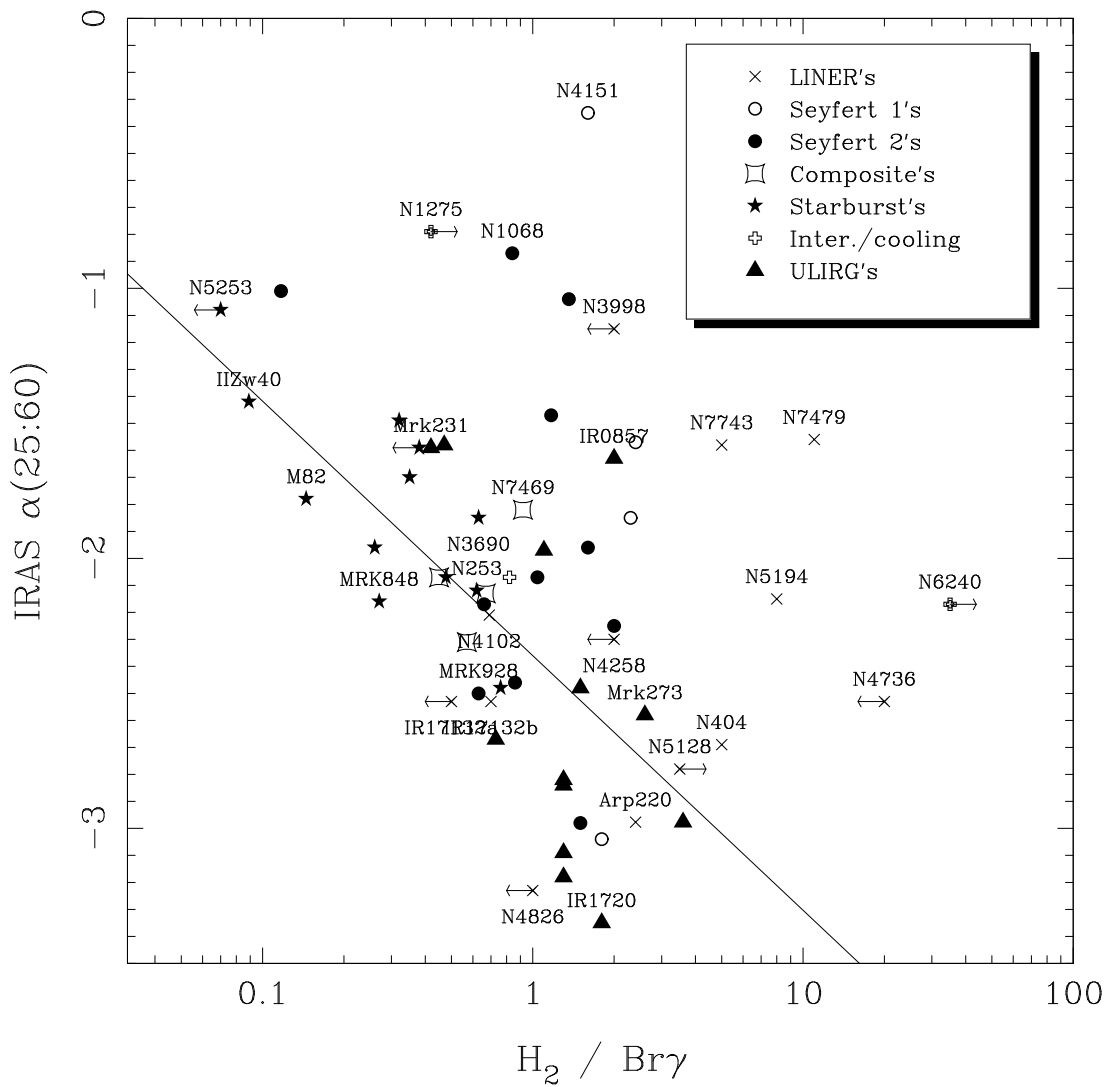}{7truein}{0}{90}{90}{-34}{-650}
\caption{ For all of the objects in this sample plus objects in
the literature, the infrared spectral index from 25 to 60 $\mu$m is
plotted against H$_2$/Br$\gamma$. A line is drawn showing the
correlation between these two parameters found for starburst galaxies
and ultraluminous IRAS galaxies in Mouri and Taniguchi (1992).  The
LINERs, like the Seyferts, do not in general follow this correlation.
Again note that Arp~220 and NGC~5128 are shown as LINERs because of
their spectral classification but they are much more energetic than the
LINERs in this sample. }
\end{figure}

% figure 7
\begin{figure}[hbtp]
\plotfiddle{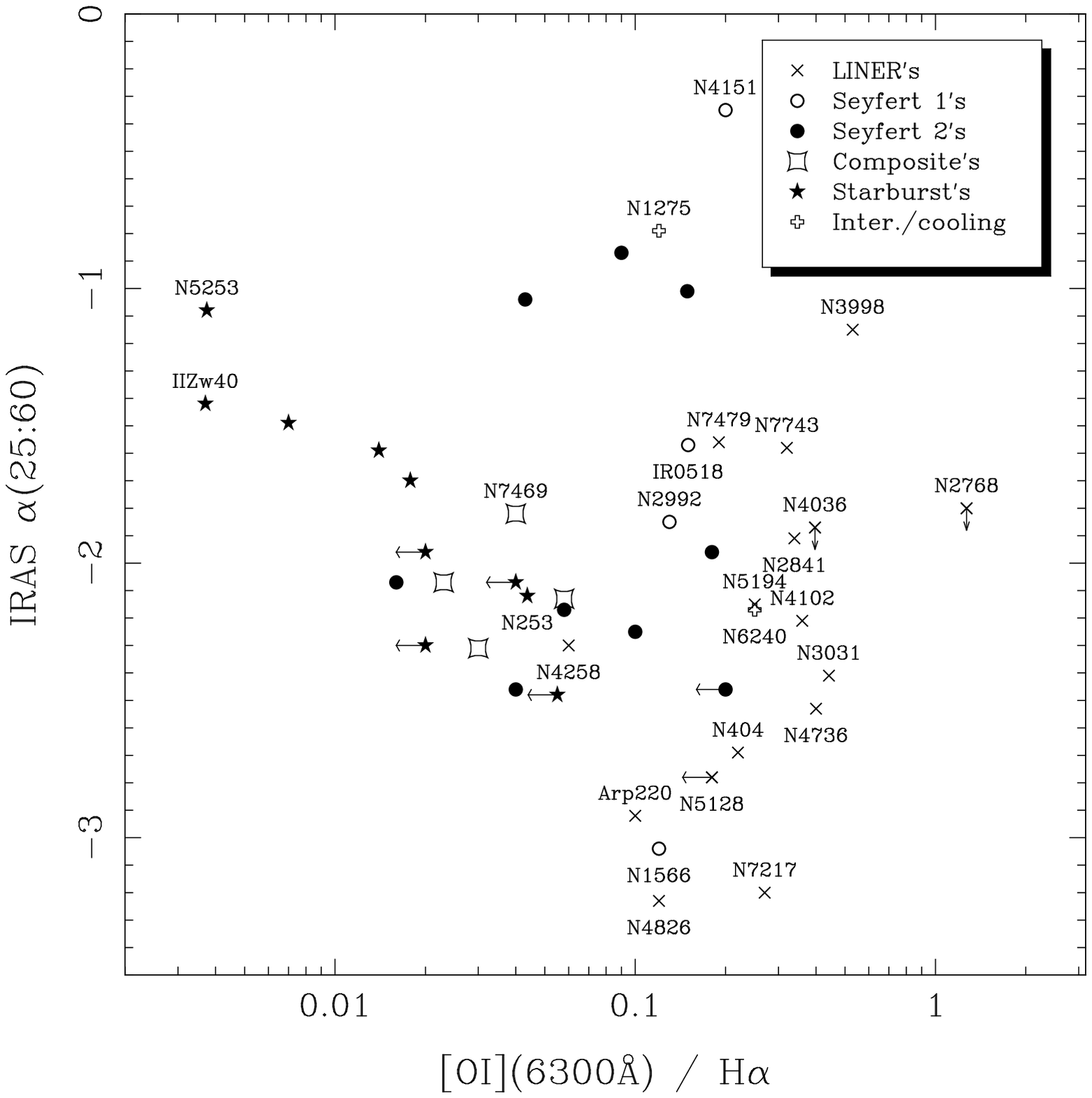}{5truein}{0}{90}{90}{-34}{-500}
\caption{ For a collection of galaxies in the literature,
the infrared spectral index from 25 to 60 $\mu$m is plotted against
[OI]/H$\alpha$. A linear correlation between these two parameters was
found for starburst galaxies in Mouri and Taniguchi (1992).  The
LINERs, like the Seyferts, do not appear consistent with this
correlation. }
\end{figure}

It is interesting to point out that the LINERs with the flattest
$\alpha$(25:60) (NGC~3998, NGC~7479, NGC~7743, and NGC~4258) and are
the furthest from the starburst correlation, have lower
[FeII]/Pa$\beta$ than the steeper spectrum LINERs.  These four objects
are also the strongest candidates in the sample to have Seyfert-like
nuclei (in fact NGC~7479 and NGC~4258 are often classified as Seyfert
2's). Also NGC~3998 is the strongest x-ray source among the LINER
sample, and has recently been found to have an unobscured hard (1-12
keV) compact, power-law, x-ray source by ASCA (Serlemitsos 1995).  For
the LINERs with steep $\alpha$(25:60) the IRAS index provides little
information except that the central source does not alter the global
infrared colors significantly.  The majority of these sources,
however, do have the highest ratios of [FeII]/Pa$\beta$.

\clearpage

\subsection{Comparison with x-ray measurements}

If the LINERs with the flattest $\alpha$(25:60), and the lowest
[FeII]/Pa$\beta$ are x-ray heated by a power-law source, as in Seyfert
galaxies, there should be sufficient x-rays to power the observed
infrared line emission.  The ratio of [FeII]/Pa$\beta$ is still
relatively high in these objects, however, and some degree of grain
destruction to increase the gas phase iron abundance is probably
necessary.  A successful x-ray heating model must therefore include a
significant hard x-ray component.  Unfortunately, very little is known
about the hard x-ray output of LINERs.  The best survey for comparison
is from the Einstein satellite which operated out to 3.5 keV.  For
Seyfert 1's the luminosity in the range 0.5 to 3.5 keV is thought to
be comparable to the hard x-ray luminosity (Van Der Werf et al. 1993).

Figure 8 plots the Einstein flux (0.2 to 3.5 keV) for the available
LINERs in the sample against the total nuclear [FeII] flux.  To obtain
the total nuclear flux, we have applied the same aperture correction
discussed in section 3.1.  This correction factor (roughly a factor of
3 for all objects) was determined by placing a synthetic aperture of
the same width as our slit on H$\alpha$+NII images (Larkin et al.
1997 in preparation) and comparing this to the total nuclear flux.  In
models of hard x-ray heating, about 1\% of the 1 to 10 keV energy can
come out in the 1.2567$\mu$m [FeII] line.  In the figure the 1\% curve
is plotted as a rough approximation of the minimum efficiency of x-ray
to [FeII] conversion.  The fact that many of the Seyfert 2's are below
this line probably results from dust obscuration of the soft x-ray
flux.  For the LINERs with Einstein detections near to or above the
1\% line, there are probably sufficient x-rays to power the iron
lines.  As in Seyfert 2's, dust obscuration may have a significant
effect on the soft x-ray fluxes of many of the LINERs making it
difficult to estimate the role of x-ray heating in those objects
without x-ray detections and those below the 1\% line.  For the five
LINERs in this sample, the ratios of x-ray(.2-3.5~KeV) flux to [FeII]
flux are: 660 for NGC~3998; 440 for NGC~4258; 70 for NGC~4736; 70 for
NGC~4826 and 175 for NGC~5194.

% figure 8
\begin{figure}[hbtp]
\plotfiddle{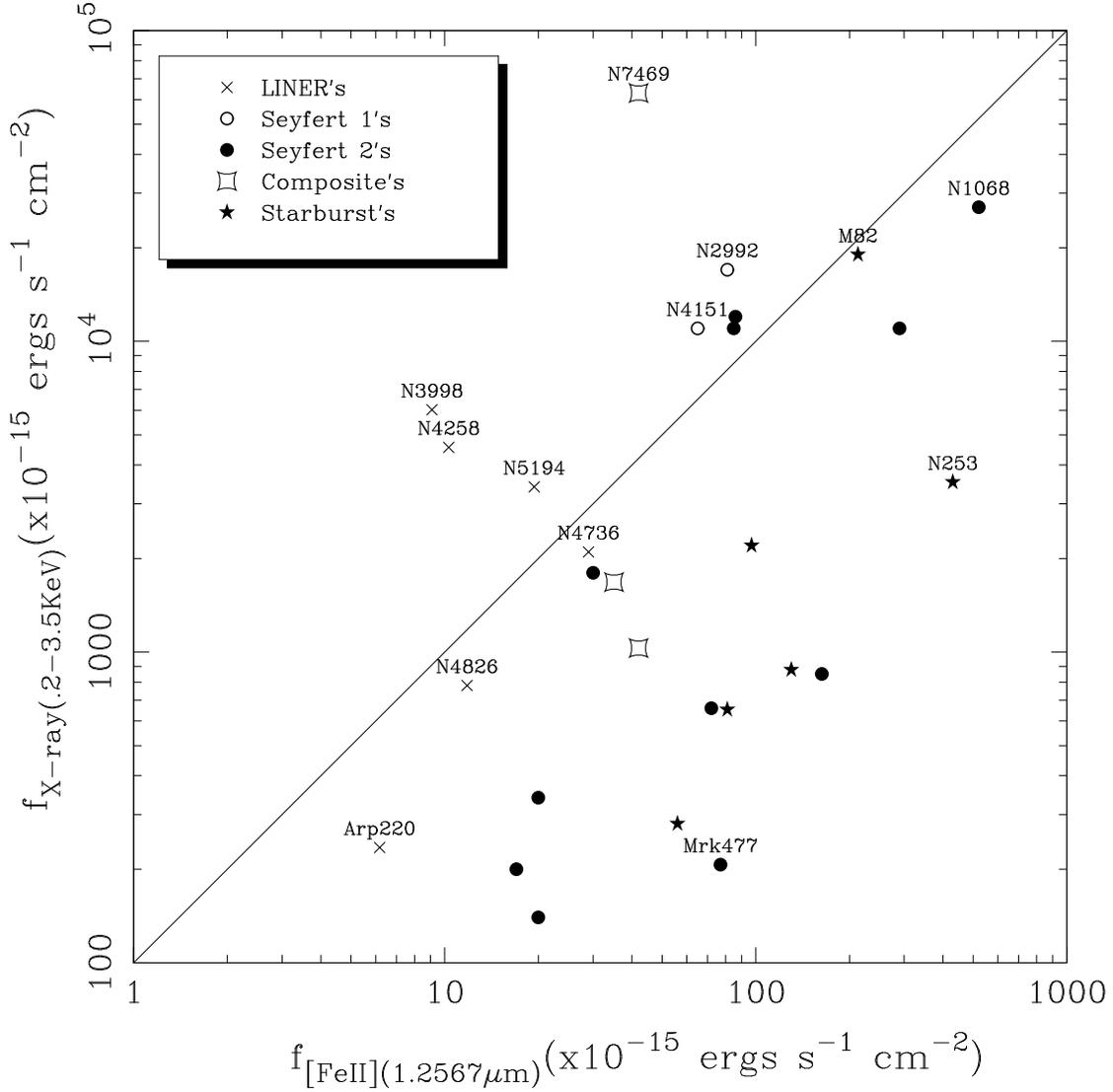}{5truein}{0}{90}{90}{-34}{-500}
\caption{For all of the objects in this sample plus objects in
the literature, the observed Einstein x-ray fluxes (.2-3.5 keV) are
plotted against the total nuclear [FeII] fluxes. A line is drawn at f$_{x-ray}$
= 100 f$_{[FeII]}$ which is approximately the fraction of [FeII] emission
expected for a given f$_{x-ray}$.  The total nuclear [FeII] flux is obtained
by applying an aperture correction to the spectroscopic measurements as
discussed in the text.}
\end{figure}

The first few LINERs (NGC~3998, NGC~4579, NGC~3031 (M81), NGC~4258,
NGC~5194 (M51)) observed with ASCA all require a power-law x-ray
source out to 12 keV, which appears to be point-like (Serlemitsos
1995).  Although the current ASCA sample is not a randomly selected
sample and the result is only preliminary, it does support x-ray
heating in at least some LINERs.

\clearpage

\subsection{Atomic Absorption Features}

In addition to the emission lines, several atomic absorption features
are seen in many of the galaxies.  These features were found by
shifting the spectra to the rest frame and noting that several of the
galaxies showed the same features.  Further confidence that they are
real features in the galaxies' spectra and are not a reduction
artifact comes from a comparison with optical spectra in Filippenko
and Sargent (1985) which show that the galaxies with the strongest
infrared absorption lines also have the strongest optical absorption
lines.  In figure 1, these suspected
absorption lines are marked by vertical marks at the top of each panel
with the most likely identification given.  The line identifications were
found by comparing the spectra with a very high resolution (R $\sim$
100,000) spectrum of the solar photosphere (Hall 1973). Among the
strongest lines present in the galaxies spectra are a Na doublet
($\lambda\lambda$2.2062,2.2090$\mu$m) and a Mg doublet
($\lambda\lambda$1.2423,1.2433$\mu$m).  The Na doublet is seen in all
seven galaxies where the spectra extend to this wavelength.  One
feature at $\lambda_{rest}$ = 1.2893$\pm$0.0003~$\mu$m is still not
identified, although it is detected in 5 or 6 of the galaxies. It is
possible that this line is from a molecular species present in cool
giants but not in the Sun.

%-----

\section{Discussion}

The most striking feature of the infrared spectra is the strength of
the [FeII] and H$_2$ features in relation to hydrogen recombination
lines, and the correlation of H$_2$ and [FeII] with the optical [OI]
line.  Ho et al. (1993) found that there was a clear separation
between LINERs and transitional objects at [OI]/H$\alpha$ $\sim$~{$1
\over 6$}.  We find a similar separation between the majority of
LINERs in this sample and Seyferts at [FeII]/Pa$\beta$ of $\sim$ 1,
and H$_2$/Br$\gamma$ ratio of $\sim$ 3. This transition is clearer
than the difference in [OI]/H$\alpha$ ratios for LINERs and Seyferts.
The one galaxy which does not satisfy this is NGC~3998, which as
argued above, satisfies the LINER criterion only for larger apertures
than employed here, and which is Seyfert-like in small apertures.  In
this object at least, the LINER classification may be based solely
upon circumnuclear emission lines which may be photoionized or shock
heated by a central AGN.

The H$_2$/Br$\gamma$ ratio also divides the Seyfert's from the
starburst galaxies in the literature at $\sim$ 0.6, although several
exceptions are evident.  It is therefore, interesting to consider that
H$_2$/Br$\gamma$ can alone be used to classify galaxies as either
starbursts, Seyferts (and transitional LINERs), and LINERs.  Figure
5 shows that by including the [OIII]/H$\beta$ ratio, the
divisions between these classes become even clearer.  It must,
however, be remembered that these trends apply only when the
true Pa$\beta$ and Br$\gamma$ emission line fluxes free of stellar
absorption are available or can be estimated with reasonable
precision.  This is particularly important for LINERs where the
stellar features often dominate the nuclear activity even for
small apertures.

\subsection{Possible Sub-Classes}

An interesting result of this work is that the LINERs with infrared
line detections break up into basically two groups based on the ratio
of [FeII] to Pa$\beta$.  In the following, LINERs with
[FeII]/Pa$\beta$ above 2 will be referred to as ``strong'' [FeII]
LINERs, while the ones below 2 will be called ``weak'' [FeII] LINERs.
The [FeII]/Pa$\beta$ line ratio uses the extrapolated Pa$\beta$ line
fluxes discussed in section 3.1.  If the extinction towards the nucleus
were very patchy it could affect these extrapolated values and therefore
the classification scheme.  Since most of the galaxies are either nearly
face on spirals or ellipticals with little dust, we don't believe this
is a large affect.  The line ratios can equally be thought of as
the ratio of [FeII] to H$\alpha$ corrected for the measured extinction
between H$\alpha$ and H$\beta$.

The ``weak'' [FeII] LINERs include NGC~4826, NGC~3998, NGC~4258,
NGC~7479 and NGC~7743.  All three of the LINERs with relatively flat
far infrared spectra ($\alpha$(25:60)$>$-2) are ``weak'' [FeII]
LINERs.  Although the interpretation of the $\alpha$(25:60) index is
complicated, a flatter spectrum probably indicates a strong nuclear
source that is able to alter the global dust temperatures and affect
the IRAS colors. The only two LINERs with detected broad H$\alpha$
lines (NGC~3998 and NGC~4258) are ``weak'' [FeII] LINERs. These two
LINERs are also the strongest x-ray sources in the sample and have
the highest ratios of x-ray flux to [FeII] flux.  The other
three ``weak'' [FeII] LINERs also show evidence for Seyfert-like
activity. Technically, NGC~7479, should actually be classified as a
Seyfert 2 based on its optical line ratios. NGC~7743 also has unusual
line ratios with the highest [OIII]/H$\beta$ ratio in the sample
indicating a relatively high state of ionization.  The last ``weak''
[FeII] LINER, NGC~4826 is technically a ``weak'' [OI] LINER since
[OI]/H$\alpha$ is less than $1 \over 6$, and it has the lowest
H$_2$/Br$\gamma$ ratio in the sample.

The ``strong'' [FeII] LINERs are NGC~404, NGC~4736, NGC~5194 and
NGC~7217.  None of the ``strong'' [FeII] LINERs has a detected broad
H$\alpha$ line, and none are known as strong x-ray sources.  Among the
``strong'' [FeII] LINERs, NGC~404 and NGC~4736 are particularly
noteworthy for having unusually strong Balmer absorption lines
indicative of recent star formation.  NGC~4736 and NGC~7217 also have
inner rings or arcs of H$\alpha$ emission probably resulting from
circumnuclear star formation.  NGC~5194 was classified by Heckman
(1980) as a transition object (its oxygen line ratios are on the
border between Seyfert and LINER classifications).  It does, however,
have a high ratio of [FeII]/Pa$\beta$ (2.3) and a very strong H$_2$
line (H$_2$/Br$\gamma$ = 8).  Because NGC~5194 is closer to a Seyfert
in optical line ratios, and because it has a good x-ray detection,
however, it is not as consistent with star formation as the other
``strong'' [FeII] LINERs.  All of the ``strong'' [FeII] LINERs have
steep ($\alpha$(25:60)$<$-2) IRAS indices.

An appealing model is that the two groupings reflect different
excitation mechanisms (See the following subsection). The ``weak''
[FeII] LINERs may be low luminosity Seyferts, while the ``strong''
[FeII] LINERs may be powered by compact starbursts. Compact supernova
remnants (pressure confined) are likely to enhance the [FeII] line in
the latter group.  The [FeII] luminosities of several of the galaxies,
in particular the ``strong'' [FeII] LINERs NGC~404 and NGC~4736, are
only a few times those expected for individual supernova remnants,
making supernovae, an attractive excitation source.  Although this
model of ``strong'' and ``weak'' [FeII] LINERs is attractive it is by
no means definitive, since none of the objects in either sample is
unambiguously a Seyfert-like or starburst-like nucleus.  The next
subsection also shows that x-ray heating from a central Seyfert-like
source can reproduce the large [FeII]/Pa$\beta$ ratios of ``strong''
[FeII] LINERs under special conditions.  It is, however, suggestive
that the LINERs with the strongest indications of recent star
formation have the greatest [FeII]/Pa$\beta$ ratios.  It is possible
that the ``strong'' [FeII] LINERs have a Seyfert-like core, but also
have a surrounding star formation region where supernova remnants
enhance the [FeII] emission.

\subsection{Excitation of the [FeII] and H$_2$ emission}

The main goal of this study was to characterize the infrared spectra of
LINER galaxies with the further goal of determining the source of the strong,
low ionization emission lines in LINER spectra.  The observation that
both the molecular line H$_2$ 1-0 S(0) and the [FeII] line are strong,
places stringent requirements on the excitation mechanism
since (1) H$_2$ is relatively easy to destroy, and (2) 98\% of the iron is
tied up in dust grains in the local ISM.  In order for a common
mechanism to power both lines, as well as the optical [OI] line, it
must not destroy all of the H$_2$ molecules and yet must free up iron
through dust grain destruction.  The strength of the [FeII] line is
much harder to explain through increased metallicity, since the
oxygen, nitrogen and sulfur lines in the optical are not consistent
with extremely high metallicities.

\vfill\eject
A general equation for the strength of [FeII] versus Pa$\beta$
is derived from Blietz et al. (1994) and is given by:

\begin{eqnarray}
\label{e:coef}
{[FeII](1.2567\mu m) \over Pa\beta} & = & coef \times \delta \times
\left( {f_{II} \over \xi} \right) \\
{\rm where} & & \nonumber \\
coef & = & {22 \times T_4^{0.07} \times exp \left( {-1.57 \over T_4}
\right) } \over { \left[ 1 + 4.2 \times \left( {T_4^{0.69} \over n_4}
\right) \right] }. \nonumber
\end{eqnarray}

\noindent
where T$_4$ is the temperature in units of 10$^4$ K, n$_4$ is the
electron density in units of 10$^4$ cm$^{-3}$, $\delta$ is the
fraction of iron in the gas phase, f$_{II}$ is the fraction of gaseous
iron that is singly ionized, and $\xi$ is the ionization fraction of
hydrogen.  The coef term is provided for convenience to separate the
temperature and density dependence from the dependences on iron
abundance and ionization states.  This equation is of course only
valid if both lines arise within the same parcel of gas.  If multiple
excitation mechanisms exist, then the equation can still be used to
try and constrain the dominant mechanism.

Since $\delta$ is $\sim$0.02 in the local ISM, even for the most
extreme possible temperature and density for these objects, coef
$\times$ $\delta$ is only 0.5.  So unless the gas phase abundance of
iron is much greater than locally, and/or the ionized fraction of iron
to hydrogen is much greater than 1, then [FeII]/Pa$\beta$ cannot reach
the level of the ``strong'' [FeII] LINERs.

Several types of environments can immediately be ruled out based on
equation \ref{e:coef}.  Standard HII regions have a fraction
of Fe$^+$ to hydrogen of less than $\sim$0.2 (Oliva et al. 1989).  In
order to reach [FeII]/Pa$\beta$ of even the values of $\sim$0.6 seen
in the ``weak'' [FeII] LINERs requires the gas phase abundance of iron
($\delta$) to be 0.12 (a six fold increase in gas phase iron) for even
the most extreme conditions (T = 10$^5$ K and n = 10$^5$ cm$^{-3}$).
For more reasonable conditions (T = 10$^4$ K and n = 10$^4$
cm$^{-3}$), $\delta$ must be greater than 1, which is unphysical.
Even $\delta$ of 0.12 is difficult to explain in an HII region, since
there are no strong mechanisms to process dust grains.  Since Galactic
HII regions are observed to have very weak [FeII] emission, it is
expected that extragalactic HII regions should have [FeII]/Pa$\beta$
ratios much less than unity.

The photodissociation region around strong UV sources were also
examined by Blietz et al. (1994) since these regions can produce
strong H$_2$ and other low ionization lines. The temperatures in the
photodissociation regions are always below 2000 K, however, which
greatly reduces the [FeII] flux (coef $<$ 1).  Also these regions tend
to be very thin, so the ratios of low ionization lines to hydrogen
recombination lines (which are generated interior to the ionization
fronts) is in general very low.

X-ray heating from a nonthermal power-law source can produce an
extended ionization front if very hard x-rays are present (Halpern and
Grindlay 1980).  In such extended regions, the fraction of ionized iron
to hydrogen can reach 100, although the temperature may be only
$\sim$3000 K (Blietz et al. 1994). For this temperature, the coef
parameter is about 0.11 for any density over $\sim$10$^4$ cm$^{-3}$.
For a ratio of [FeII]/Pa$\beta$ as high as 2, $\delta$ must be at
least 18\%.  Although high, this is not unfeasible since hard
x-rays can destroy dust grains.  For the ``weak'' [FeII] LINERs, the
ratio of [FeII]/Pa$\beta$ of $\sim$0.6, $\delta$ need only be 0.055,
which is very reasonable.  X-rays are therefore an attractive emission
source for the ``weak'' [FeII] LINERs and may play a part in powering
the infrared emission lines in the ``strong'' [FeII] LINERs as well.

Another possible excitation mechanism is a J-type shock (jump
discontinuity shocks without magnetic precursors).  Shocks can produce
large fractions of ionized iron to hydrogen and provide a natural
mechanism for dust grain processing.  Seab and Shull (1983) find that
fast shocks (40 to 120 km s$^{-1}$) can increase the gas phase iron
abundance ($\delta$) up to 60\%.  For an expected ionization fraction
of iron to hydrogen of around 2, J-shocks then require coef to be
greater than 1.7 for [FeII]/Pa$\beta$ around 2, and 0.5 for
[FeII]/Pa$\beta$ of 0.6.  These values are possible for virtually any
temperature above 10$^4$ and density above 10$^4$.  This fact makes
shocks a very attractive mechanism for the ``strong'' [FeII] LINERs,
and they may also produce the [FeII] emission in the ``weak'' [FeII]
LINERs. If present in ``weak'' [FeII] LINERs, the effect of shocks is
probably diluted by other mechanisms.  Although fast shocks will
destroy H$_2$, there may exist a range of gas densities making a
variety of shock speeds possible.  The strong H$_2$ lines may then
result from the same shock mechanism that produces the [FeII]
emission, but in different parts of the gas.

Both hard x-ray heating, and fast shocks are plausible mechanisms for
all of the LINERs, although the ``weak'' [FeII] LINERs are easier to
explain by x-ray heating while ``strong'' [FeII] LINERs are more
consistent with shock excitation.  Both mechanisms are also able to
generate the H$_2$ emission, but shocks are more efficient at [FeII]
production, and are better able to explain the ``strong'' [FeII]
LINERs. If shocks are present, either compact supernova remnants
(pressure confined), or other mechanisms such as cloud-cloud
collisions, or outflowing winds are plausible shock sources. Supernova
remnants are the likely source for the ``strong'' [FeII] LINERs since
many of these show other evidence for recent star formation.

\vfill\eject
\subsection{Possible Biases}

In any survey of a small sample of objects with faint emission
features, biases in selection, photometry and aperture correction,
sensitivity and a large number of other sources may play a significant
role. In this subsection we discuss some of the most obvious sources
of bias, and try to estimate their effects.

One of the largest biases affecting this survey is the size of the
sample and the fact that the line detections are preferentially in the
most active LINERs; the others are too faint.  A significant
population of LINERs have featureless infrared spectra and cannot be
addressed by the infrared diagnostics.  For most quantities discussed,
there is a relatively smooth transition from LINER-like spectral
properties to Seyfert-like.  An important consideration is that the
least active LINERs, which have no line information, are less like
Seyfert's in their emission mechanism than those discussed in detail,
and many of the correlations may not hold for these LINERs.  We have
tried, in a limited fashion, to address this problem by also using the
optical [OI](6300\AA) line, which is well correlated with the infrared
lines for all LINERs, in examining the excitation mechanisms.  Since
this line must be detected for a galaxy to be classified as a LINER
according to the original definition, it does not suffer from the same
selection bias. However, it is possible that the [OI] emission does
not arise from the same regions as the infrared lines.

Another bias is the lack of hydrogen recombination line detections.
Since we are applying optical extinction estimates to extrapolate the
intrinsic strengths of the near infrared features, we may be subject
to large uncertainties if the extinctions to the optical and IR
emitting regions are significantly different.  If the extinction to
the infrared lines were higher than predicted from H$\alpha$ and
H$\beta$, then the estimated infrared recombination lines would be too
low, and the corresponding ratios with [FeII] and H$_2$ too
high. Since the extrapolated [FeII]/Pa$\beta$ and H$_2$/Br$\gamma$
line ratios correlate with the [OI] line, it is not likely that the
recombination strengths have been severely underestimated.  Also,
since the extrapolated values of Pa$\beta$ and Br$\gamma$ from optical
lines are typically higher than can be accounted for by correcting the
observed spectra for stellar absorption, it is more likely that
Pa$\beta$ and Br$\gamma$ have been overestimated which would imply
that the line ratios are even higher than calculated.  A logical next
step will be to obtain high signal to noise infrared spectra of
several template galaxies for subtraction in order to identify
extremely weak emission lines lost in stellar absorption features.

It is possible that the difference in size between our aperture and
those used in the optical plays a role in increasing the observed
[FeII] and H$_2$ ratios.  Knop (personal communication 1997) has used
a similar small aperture to measure the infrared line ratios in some
nearby Seyfert galaxies and has found that many of them have slightly
higher [FeII] and H$_2$ flux ratios than Seyferts in the literature,
but the effect is much too weak to account for the differences seen
between the LINERs and the Seyferts.  There is one object (NGC~2110)
in the Knop sample, however, which is a Seyfert with a very large
[FeII] to Pa$\beta$ flux ratio of 8, and an H$_2$ to Br$\gamma$ flux
ratio of 3.  This object is extremely interesting but does not reflect
the ratios typically observed for Seyfert galaxies in small apertures.
In fact, several of the LINERs that have multi-aperture optical
spectra, show stronger low ionization lines in the larger aperture,
which implies just the opposite effect.

In general, we believe that the biases and observational limitations
present in the data do not alter the interpretation of the physical
processes at work in LINER nuclei.

%-----------------------------------------------------------------

\section{Summary and Conclusions}

This paper has described an infrared spectroscopic survey of 12
``classical'' LINER galaxies.  The spectra have concentrated on the
[FeII](1.2567$\mu$m), Pa$\beta$, H$_2$(2.1218$\mu$m) and Br$\gamma$
infrared lines.  The major results are:

\begin{enumerate}

\item {[FeII] and H$_2$ are the strongest infrared lines in classical
LINERs.  Using extrapolated H$^+$ line strengths from the optical,
approximately half of the classical LINERs have ratios of
[FeII]/Pa$\beta$ and/or H$_2$/Br$\gamma$ a factor of two or more
higher than typical Seyfert galaxies and a factor of five or more
higher than typical starburst galaxies.}

\item {A natural subdivision between the LINERs occurs at [FeII]/Pa$\beta$
= 2. The four ``strong'' [FeII] LINERs exhibit evidence for recent or
ongoing star formation.  As a group, the [FeII] emission in these
LINERs is consistent with shock excitation from compact supernova
remnants.  The five ``weak'' [FeII] LINERs have more in common with
Seyfert galaxies, including the only two objects in the sample with broad
H$\alpha$ (NGC~3998 and NGC~4258). The lower [FeII]/Pa$\beta$ and
H$_2$/Br$\gamma$ ratios of ``weak'' [FeII]
LINERs are consistent with hard x-ray heating from a power-law
source.  Neither excitation mechanism is ruled out for either type of
object, however, and it is possible that it is just the relative
strengths of the two mechanisms that are different in the two groups.}

\item {In most of the LINERs, the estimated amount of Pa$\beta$
absorption from the stellar population can account for the lack of
Pa$\beta$ detected in emission. Several of the LINERs have unusually
strong Pa$\beta$ absorption ( $>$2$\AA$ EQW) indicative of younger
stellar populations.  The five galaxies with the strongest absorption
all have strong [FeII] lines and supernova remnants from recent star
formation are a plausible explanation for their enhanced [FeII]
strengths.}

\item {A strong linear correlation is observed between
H$_2$/Br$\gamma$, [FeII]/Pa$\beta$ and [OI]/H$\alpha$ for a range of
100 in all ratios and for all of the observed galaxy types.  However,
the ``strong'' [FeII] LINERs are stronger in [FeII] than the
correlation with [OI] predicts.  Both the H$_2$/Br$\gamma$ and to a
lesser degree the [FeII]/Pa$\beta$ ratio, appear to separate the three
galaxy classes: LINERs, Seyferts and starbursts. Only six objects have
been observed to have H$_2$/Br$\gamma$ higher than 3, and all are
LINERs.  In conjunction with the [OIII]/H$\beta$ ratio, the class
distinctions are even clearer.}

\item{The shallow far infrared spectral slopes of some of the ``weak''
[FeII] LINERs in comparison to the strengths of H$_2$ and [OI], appear
inconsistent with the correlations observed for starburst and ULIRG
galaxies.  This argues very strongly that a nonthermal heat source is
providing much of the dust heating and line excitation for these
objects.}

\item {LINERs with x-ray detections appear to have sufficient x-ray
luminosity to power the observed infrared lines.  Starbursts often
have [FeII] and H$_2$ lines many times stronger than the x-rays
luminosity should be able to produce. Again the strong x-ray LINERs do
not appear consistent with star formation, but instead behave like
mini-Seyferts.}

\end{enumerate}

\acknowledgments

The authors would like to thank Thomas Murphy, Gerry Neugebauer, David
Shupe, and Alycia Wienberger for many useful discussions on these
results.  We also want to thank Roger Blandford and Nick Scoville for
many helpful comments.  Special thanks goes to the Palomar night
assistants, Skip Staples, Juan Carasco and the entire Palomar staff.
This research has made use of the NASA/IPAC Extragalactic Database
(NED) which is operated by the Jet Propulsion Laboratory, Caltech,
under contract with NASA.

\vfill\eject
%------------------------------------------------
%------------------------------------------------
\appendix{Individual Objects}

\sl NGC~404 \rm: A classical LINER fully satisfying the Heckman (1980)
criteria.  The infrared spectra show very prominent [FeII] emission
and a strong detection of H$_2$.  NGC~404 has the largest [FeII] line
to continuum ratio for any of the LINERs in this sample.  It is
plausible that the estimated Pa$\beta$ flux is too high and that the
calculated [FeII]/Pa$\beta$ flux is significantly higher than the
value of 2.7 in table 3.  Even with this large estimated Pa$\beta$
flux, NGC~404 has one of the highest ratios of [FeII] to Pa$\beta$ of
any extragalactic object.  The apparent feature at the location of
Br$\gamma$ is due to residual ripple in the G star spectrum used to
remove telluric absorption which is most apparent in NGC~404 because
it is at essentially zero redshift.  Since the expected Br$\gamma$ is
below the detection limit, and there is an obvious explanation for the
feature, we do not regard this as a real detection of Br$\gamma$ in
NGC~404. All of the line emission is spatially unresolved with a
seeing of 1''.  At a distance between 1.8 and 2.4 Mpc, this
corresponds to only 10pc. At this distance, the aperture corrected
[FeII] luminosity is 1.1x10$^{37}$ ergs s$^{-1}$ which is only a few
times the expected luminosity for a single supernova remnant (Moorwood
and Oliva 1988). It is an interesting possibility that the emission
within NGC~404 could be generated in a single SNR in a high density
medium.  The Pa$\beta$ absorption correction, estimated in subsection
3.2 is about a factor of two too low to explain why Pa$\beta$ is
undetected.  The most obvious explanation is that NGC~404 contains a
significant population of young stars which increase the amount of
Balmer absorption. This is supported by Keel (1983), who considered
NGC~404 a peculiar member of the LINER class due to the presence of a
blue continuum and strong Balmer absorption lines.  NGC~404 shows none
of the infrared atomic absorption features seen in some of the other
galaxies in this sample.  The optical spectrum, (Filippenko and
Sargent, 1985) also shows a relatively smooth continuum spectrum.
Although the emission lines are very narrow, a strong central UV point
source and several surrounding fainter UV sources have been detected
with HST (Maoz, 1993).  Although the central UV source was taken as
evidence for an AGN by Maoz, a central star cluster is also a
possibility.

%-----

\sl NGC~2685 \rm: Classified as a LINER based upon only the [OII]/[OIII]
ratio, this object has very weak optical lines.  No emission lines are
detected in the infrared spectrum.  There appears to be a strong
Pa$\beta$ absorption feature, although the continuum around it has
residual features from the night sky and the G star division.  It does
show most of the atomic absorption features, especially the Mg
($\lambda\lambda$1.2423,1.2433$\mu$m) and Na
($\lambda\lambda$2.2062,2.2090$\mu$m) lines.  Optically it has strong
Balmer absorption, weak NII and strong atomic absorption lines
(Filippenko and Sargent 1985).

%-----

\sl NGC~3992 (M~109) \rm: Like NGC~2685, it was classified as a LINER
solely on the basis of the [OII]/[OIII] ratio.  No emission lines are
detected in the infrared spectrum.  There is an obvious strong
Pa$\beta$ absorption line. NGC~3992 also shows the strongest atomic
absorption features of the classical LINERs in this sample. The
spectrum in Filippenko and Sargent (1985) also shows prominent atomic
absorption features and weak emission lines.

%-----

\sl NGC~3998 \rm: A true LINER which shows increasing ratios of [OI]
and [OII] to [OIII] with increasing aperture size.  The only detected
infrared emission line is [FeII] which is detected at the 4 to 5 sigma
level.  Using the estimated Pa$\beta$ flux, the ratio of [FeII] to
Pa$\beta$ is 0.6, which is one of the lowest in the sample.  H$_2$ is
undetected, setting a low limit for the H$_2$/Br$\gamma$ ratio.  Both
of these ratios are significantly lower in comparison to the
OI/H$\alpha$ ratio than the other LINERs in the sample.  The limit of
the ratio of H$_2$ to [FeII], however, does lie along the main
correlation for LINERs and other galaxies.  In Larkin et al. (1997, in
preparation), it is argued that the NII emission is significantly
extended in comparison to the H$\alpha$.  It is probable that the OI,
and also probably the [FeII] and H$_2$, are also more extended than
the H recombination lines, and that the optical ratio calculated in
the larger aperture is enhanced in comparison to the infrared ratios
in a much smaller aperture.  If true, then the infrared ratios in a
larger aperture would probably be greater and NGC~3998 would lie
closer to the main correlations shown in figures 2 and
3.  In principle, this says that the low ionization lines
are more extended than H$\alpha$, and that the unresolved core has
more Seyfert-like line ratios.  The Na and Mg absorption features may
be present at a very low level.  Optical spectra show prominent broad
H$\alpha$ wings and even a broadened [OI] line (700 km sec$^{-1}$).
Optical absorption features, if present, are weak and broadened
(Filippenko and Sargent 1985).  Like NGC~404, NGC~3998 has had a
strong UV point source detected by HST (Fabbiano et al. 1994). Coupled
with the broad H$\alpha$, this makes NGC~3998 a prime candidate for a
low luminosity Seyfert galaxy.

%-----

\sl NGC~4258 (M~106) \rm: One of Heckman's original two transitional
objects.  Its optical line ratios are actually typical of Seyferts.
Its infrared spectra are similar to those of NGC~3998; only a fairly
weak detection of [FeII] in emission.  It also has a ratio of [FeII]
to Pa$\beta$ ratio of 0.6, which is low for this sample, but similar
to NGC~3998.  Optically, is also very similar to NGC~3998, with faint
broad wings of H$\alpha$ and slightly broadened OI.  The Na absorption
feature is detected weakly. Unlike NGC~3998, NGC~4258 shows several
strong optical absorption features (Filippenko and Sargent 1985).  A
wealth of high resolution observations have focused on NGC~4258,
showing it to have a rotating ring of water masers (e.g. Nakai et
al. 1993), optical jets (e.g. Ford et al. 1986) and a steeply rising
velocity curve (Miyoshi et al. 1995), all of which make NGC~4258 one
of the best candidates for harboring a central black hole. Its [FeII]
and H$_2$ ratios are comparable to Seyferts.

%-----

\sl NGC~4589 \rm: Classified as a LINER based on only the [OII] to
[OIII] ratio.  This is the most featureless of the LINERs in the
infrared, showing no emission or absorption lines, except possibly
weak Pa$\beta$ absorption.  Mollenhoff and Bender (1989) showed that
NGC~4589 had a peculiar velocity field with evidence for large
streaming motions, and a warped dust lane. They argued that NGC~4589
was the result of a recent merger.

%-----

\sl NGC~4736 \rm: A true LINER satisfying both of Heckman's (1980)
criteria.  NGC~4736 is one of the strongest continuum sources in the
sample making the weak lines difficult to detect.  The [FeII] to
Pa$\beta$ ratio is the highest of any object in this sample, and the
second largest of any extragalactic object measured (only NGC~6240 has
a higher ratio).  This ratio is, however, uncertain by 50\% due to the
strong continuum surrounding the [FeII] line and the strong Pa$\beta$
absorption. The ratio is supported by the strong optical ratio of
[OI]/H$\alpha$ which, as discussed above, is correlated with
[FeII]/Pa$\beta$.  The H$_2$/Br$\gamma$ is only a very poor upper
limit, but based on the other two line ratios, is probably also very
high ($\sim$10).  Like NGC~404, NGC~4736 is very close (4.3 Mpc)
making the [FeII] luminosity only a few times higher than an
individual SNR.  The estimated Pa$\beta$ absorption in table 3 is not
sufficient to reconcile optical line strengths and the observed
Pa$\beta$ flux.  This may indicate the presence of a younger stellar
population.  Atomic absorption features are strong in the infrared
spectra, and in the optical spectra of Filippenko and Sargent (1985).
NGC~4736 was found to have two compact UV sources separated by 2.5''
in its nucleus by HST (Maoz et al. 1993). It also had UV arcs at 2'',
4'' and 6'' suggestive of bow shocks.  At much larger scales,
NGC~4736, has a strong ring of H$\alpha$ emission ( Larkin et
al. 1997, in preparation).  All of these features suggest that
NGC~4736 has experienced recent star formation.

%-----

\sl NGC~4826 (M~64)\rm: Probably a true LINER since [OI]/[OIII] = 0.65
(Keel 1983), but [OII] is not measured. The infrared spectra look
very similar to NGC~4736, with the relatively weak [FeII] detection,
and fairly strong atomic absorption features.  A strong Pa$\beta$
absorption line is also apparent.  Optical spectra from Filippenko and
Sargent (1985) show strong emission lines, many atomic absorption
features, but no broad H$\alpha$.  Along with NGC~4258 and NGC~3998,
it has the lowest ratios of [FeII] and H$_2$ to hydrogen lines of this
sample, and their ratios are actually located within the range typical
of Seyferts.

%-----

\sl NGC~5194 (M~51) \rm:  One of Heckman's (1980) original transition
objects with oxygen line ratios comparable to Seyferts. Among the most
famous LINER galaxies, M~51 is one of only two in this sample to show
Pa$\beta$ in emission (although it is very weak).  M~51 also shows
strong [FeII] and H$_2$ emission and has estimated ratios among the
highest in this data set, significantly above any Seyfert or starburst
galaxy.  A bubble of H$\alpha$ emission is seen just off of the
nucleus (Larkin et al. 1997, in preparation).  Known as the whirlpool
galaxy, it has an obvious interaction with NGC~5195 which may
contribute to any nuclear activity.  No broad lines are detected by
Filippenko and Sargent (1985). The only obvious atomic absorption
feature in M51 is the Na doublet, although Mg may also be present.

%-----

\sl NGC~7217 \rm:  A true LINER satisfying both of Heckman's criteria.
Only a J-band spectra is available for NGC~7217.  It shows strong
[FeII] in emission, and Pa$\beta$ in absorption, along with faint
atomic absorption features.  It ranks among the top four LINERs in
[FeII]/Pa$\beta$ ratio. In H$\alpha$, an inner ring of radius 10'' is
visible suggesting some form of recent tidal interaction or star
formation (Larkin et al. 1997, in preparation).  Filippenko and
Sargent (1985) found no broad H$\alpha$ component and only weak atomic
absorption.

%-----

\sl NGC~7479 \rm: NGC~7479 does not meet the full LINER classification
of Heckman (1980) in the small aperture of Ho et al. (1993) since
[OI]/[OIII] is below 1/3rd, but it has [OI]/[OIII] of $\sim {2\over
3}$ in Keel's larger 6'' aperture.  It is also often classified as a
Seyfert 2 in the literature.  NGC~7479 is the only galaxy in this
sample to be detected in H$_2$ but not in [FeII].  It also has a
strong Pa$\beta$ emission line detected.  The small features near the
wavelength of Br$\gamma$ are probably residual fringes from the
calibration G star spectrum and are not believed to be line
detections.  Filippenko and Sargent (1985) found no broad H$\alpha$,
although many of the lines showed a complex velocity field which
suggests a fuel supply of molecular gas.  Since the [OI] may be more
extended than the [OIII] emission, the central source may have
Seyfert-like line ratios.

%-----

\sl NGC~7743 \rm: It has a [OI]/[OIII] ratio of 0.14 which is below
the cutoff of $1\over 3$ of Heckman (1980) but it has strong [OII]
emission. The infrared spectra show strong [FeII] and H$_2$ emission
lines, and perhaps broadened Pa$\beta$ and Mg absorption lines.  Its
[FeII] and H$_2$ ratios are significantly stronger than Seyfert and
starburst galaxies.  Filippenko and Sargent (1985) found no broad
H$\alpha$ component and a fairly smooth continuum with little atomic
absorption.

%-----------------------------------------------------------------
%-----------------------------------------------------------------

\clearpage
%-----------------------------------------------------------------
%-----------------------------------------------------------------

%-----------------------------------------------------------------
%-----------------------------------------------------------------

\end{document}